\RequirePackage{lineno}
\documentclass[twocolumn,showpacs,superscriptaddress,amsmath,amssymb,nofootinbib]{revtex4-2}
\usepackage[x11names]{xcolor}
\usepackage{caption}
\usepackage[colorlinks, linkcolor=blue, citecolor=OliveGreen, urlcolor=magenta]{hyperref}
\usepackage{orcidlink}

\allowdisplaybreaks[4]
\usepackage{graphicx}
\usepackage{subfigure}
\usepackage{subcaption}
\usepackage{epsfig}
\usepackage{overpic}
\usepackage{dcolumn}
\usepackage{ulem}
\usepackage{bm}
\usepackage{color}
\usepackage{lineno}
\usepackage{xspace}
\usepackage{multirow}
\usepackage{epstopdf}
\usepackage{xcolor}
\usepackage{soul}
\usepackage{verbatim}
\usepackage{enumitem}
\usepackage{todonotes}
\usepackage{slashed}
\usepackage{cancel}
\usepackage{array}
\usepackage[toc,page]{appendix}

\definecolor{OliveGreen}{rgb}{0.4, 0.8, 0.1}

\usepackage{diagbox}

\newcommand{\moe}{\affiliation{Key Laboratory of Atomic and Subatomic Structure and 
Quantum Control (MOE), Guangdong-Hong Kong Joint Laboratory of Quantum Matter, Guangzhou 510006, China}}

\newcommand{\iqm}{\affiliation{State Key Laboratory of Nuclear Physics and 
Technology, Institute of Quantum Matter, South China Normal 
University, Guangzhou 510006, China}}

\newcommand{\gbrce}{\affiliation{Guangdong Basic Research Center of Excellence for 
Structure and Fundamental Interactions of Matter, Guangdong Provincial Key Laboratory of Nuclear Science, Guangzhou 510006, China}}

\newcommand{\hiskp}{\affiliation{Helmholtz-Institute f\"ur Strahlen- und Kernphysik (Theorie) and Bethe Center for Theoretical Physics, Universit\"at Bonn, D-53115 Bonn, Germany}}

\newcommand{\ias}{\affiliation{Institute for Advanced Simulation (IAS-4), D-52425 J\"ulich, Germany}}

\newcommand{\peng}{\affiliation{Peng Huanwu Collaborative Center for Reserch and Education, International Institute for Interdisciplinary and Frontiers, Beihang University, Beijing 100191, China}}

\newcommand{\scnt}{\affiliation{Southern Center for Nuclear-Science Theory (SCNT), Institute of Modern Physics, Chinese Academy of Sciences, Huizhou 516000, Guangdong Province, China}}

\newcommand{\OU}{\affiliation{Research Center for Nuclear Physics (RCNP), Osaka University, Ibaraki 567-0047, Japan}}

\newcommand{\uestc}{\affiliation{School of Physics, University of Electronic Science and Technology of China, Chengdu 611731, China}}

\begin{document}
\include{def-com}
\title{\boldmath Resonance parameters of the vector charmoniumlike state $G(3900)$}

\author {\mbox{Quanxing Ye}\orcidlink{0009-0006-8322-3163}}
\iqm
\moe
\gbrce

\author {\mbox{Zhenyu Zhang\orcidlink{0000-0001-7042-2311}}}
\iqm
\moe
\gbrce

\author {\mbox{Meng-Lin Du\orcidlink{0000-0002-7504-3107}}}
\email{du.ml@uestc.edu.cn}
\uestc

\author {\mbox{Ulf-G. Mei{\ss}ner\orcidlink{0000-0003-1254-442X}}}
\email{meissner@hiskp.uni-bonn.de}
\hiskp 
\ias 
\peng

\author {\mbox{Peng-Yu Niu\orcidlink{0000-0001-8455-9570}}}
\email{niupy@m.scnu.edu.cn}
\iqm
\gbrce

\author {\mbox{Qian Wang\orcidlink{0000-0002-2447-111X}}}
\email{qianwang@m.scnu.edu.cn, corresponding author}
\iqm
\gbrce
\scnt
\OU

\date{\today}
\begin{abstract}
Motivated by the updated analysis of the $G(3900)$ by the  BESIII collaboration, we perform a global analysis of the cross sections of the $e^+e^-\to D\bar{D}$, $e^+e^-\to D\bar{D}^*+c.c.$, $e^+e^-\to D^*\bar{D}^*$ processes, especially focusing on the properties of the $G(3900)$. As the energy region of interest is limited by the next opening threshold, i.e. the $D_1\bar{D}$ threshold, we focus on the energy region $[3.7,4.25]~\mathrm{GeV}$, where three charmonia $\psi(1D)$, $\psi(3S)$ and $\psi(2D)$ explicitly contribute to the cross sections. By constructing the $P$-wave contact interaction between the $(D,D^*)$ doublet and its antiparticle in the heavy quark limit, we extract the physical scattering amplitude by solving the Lippmann-Schwinger equation. No matter whether three or two charmonium states are included in our framework, we always find a dynamically generated state corresponding to the $G(3900)$, which suggests it to be a $P$-wave dynamically generated state. We also predict several dynamically generated states in the corresponding $1^{-+}$ channel. These states can be further searched for in the electron-positron annihilation process involving the emission of a single photon. 
\end{abstract}

\maketitle

\section{INTRODUCTION}

Electron-positron annihilation is one of the most important processes for shedding light on the dynamics of the strong interaction. For instance, the number of colors can be extracted from the ratio between the cross section of the $e^+e^-\to \mathrm{hadrons}$ process and that of the pure electromagnetic process $e^+e^-\to \mu^+\mu^-$. Among the former cross section, the open-charmed channels (either two-body final states or many-body final states) take up the largest fraction. The Belle~\cite{Belle:2007qxm}, CLEO~\cite{CLEO:2008ojp} and BaBar~\cite{BaBar:2006qlj} collaborations have measured the cross sections of a pair of open charmed mesons. Recently, the BESIII collaboration measured the cross sections of two-body~\cite{BESIII:2024ths,BESIII:2021yvc,BESIII:2023wsc}, three-body~\cite{BESIII:2018iea,BESIII:2023cmv}, and four-body~\cite{BESIII:2022quc} open charmed processes more precisely. As the electron and the positron annihilate into a virtual photon, this kind of process is also the most important platform for studying the normal vector charmonia and exotic vector charmonium-like states. For instance, these bring us to an opportunity to study the non-$D\bar{D}$ decay width of the $\psi(3770)$~\cite{Zhang:2009kr,Liu:2009dr,Hanhart:2023fud,Shamov:2016mxe} and the time-like electromagnetic $D^*\to D$ transition form factor~\cite{Zhang:2010zv}. Especially, the $e^+e^-\to D\bar{D}$ process provides the most precise determination of the resonance parameters of the $\psi(3770)$~\cite{Zhang:2009gy,Chen:2012qq,Cao:2014qna,Coito:2017ppc,Uglov:2016orr,Shamov:2016mxe,Husken:2024hmi,Hanhart:2023fud}. For vector charmonium-like states, the $Y(4230)/Y(4260)$~\cite{BaBar:2005hhc,CLEO:2006ike,Belle:2007dxy}, $Y(4360)$~\cite{BaBar:2006ait,Belle:2007umv}, $Y(4660)$~\cite{Belle:2007umv,BaBar:2012hpr} are measured in electron-positron annihilation process with different final states.  

Approximately twenty years ago, the Belle~\cite{Belle:2007qxm} and BaBar~\cite{BaBar:2006qlj,BaBar:2008drv} collaborations measured the cross section for the $e^+e^- \rightarrow D\bar{D}$ process. They both observed a peak structure around $\sqrt{s}=3.9$~GeV. In these works, the peak is not associated to a resonance. The BESIII Collaboration recently performed a precise measurement of the Born cross sections for the $e^+e^-\rightarrow D\bar{D}$ process~\cite{BESIII:2024ths}, which is consistent with previous results from BaBar and Belle. Apart from the $1^{--}$ charmonia $\psi(3770)$, $\psi(4040)$, $\psi(4160)$, $\psi(4360)$, $\psi(4415)$ and $\psi(4660)$, they also observed a peak structure around $3.9$~GeV. Its mass and width in the Breit-Wigner formalism are $3872.5\pm 14.2\pm 3.0$~MeV and $179.7\pm14.1\pm7.0$~MeV, respectively. Although the coupled-channel analysis of the Belle and BESIII data could produce a peak structure around $3.9~\mathrm{GeV}$ without requiring an additional new state, accurately describing the nearby points appears to be highly challenging~\cite{Uglov:2016orr,Salnikov:2024wah}. In Ref.~\cite{Zhang:2009gy}, a perturbative treatment of $\psi(2S)-\psi(1D)$ mixing is carried out within an effective Lagrangian approach, where the authors interpret the $G(3900)$ as a resonance but also demonstrate that it can be explained by the $D^*\bar{D}$ threshold. Cao and Lenske have analyzed the line shape of $\psi(3770)$ using a coupled-channel $T$-matrix approach and achieved a good fit to the experimental data, suggesting that the broad structure $G(3900)$ results from the distortion of the $\psi(3770)$ tail caused by the $D\bar{D}^*$ threshold~\cite{Cao:2014qna}. The $K$-matrix formalism is used to systematically study $e^+e^-\rightarrow D^{(*)}\bar{D}^{(*)}$ and $e^+e^-\rightarrow \text{everything}$ in Ref.~\cite{Husken:2024hmi}. The study indicates that no additional bare pole is needed to explain the data near $3.9~\mathrm{GeV}$. In the scenario of the one-boson-exchange (OBE) model, Lin {\it et al.} ~\cite{Lin:2024qcq} and Chen {\it et al.}~\cite{Chen:2025gxe} show that the existence of the $S$-wave $X(3872)$, $T_{cc}(3875)$, $Z_c(3900)$ hadronic molecules indicate the existence of a $P$-wave $D\bar{D}^*/\bar{D}D^*$ molecule state, identified as the $G(3900)$. Ref.~\cite{Du:2016qcr} also assigns the $G(3900)$ to a $P$-wave $D\bar{D}^*/\bar{D}D^*$ resonance by the contact interactions within the heavy quark spin symmetry framework, with explicit inclusion of $S$-channel charmonia contribution. Ref.~\cite{Nakamura:2023obk} also obtains the same conclusion by overall fitting to the lineshapes in various channels as well as the invariant distributions of their subsystems.
Although the later references suggest that the $G(3900)$ can be accepted as the $P$-wave $D\bar{D}^*$ resonance, it does not answer the question whether the $G(3900)$ is a dynamically generated state or a renormalized bare charmonium state. Another question is whether the existence of the $G(3900)$ is model-dependent or not. 

To answer the above questions, we construct the contact potential for the $P$-wave scattering between the $(D,D^*)$ doublet and its antiparticle in the heavy quark limit and extract the scattering amplitudes of the $e^+e^-\to D\bar{D}$, $e^+e^-\to D^*\bar{D}+c.c.$, $e^+e^-\to D^*\bar{D}^*$ by solving Lippmann-Schwinger equation (Sec.~\ref{sec:formalism}). The numerical results and discussions follow as Sec.~\ref{sec:results}. The summary and outlook is given in Sec.~\ref{sec:summary}).

\section{FORMALISM}
\label{sec:formalism}
The formalism of this work is an SU(3) extension of that in Ref.~\cite{Du:2016qcr}, in which SU(2) flavor symmetry is adopted. Considering the recent progresses from the experimental side, i.e. the measurements of the $e^+e^-\to D_s^+D_s^-$~\cite{BESIII:2024zdh,BaBar:2010plp} and $e^+e^-\to D_s^{*+}D_s^{*-}$~\cite{BESIII:2023wsc} 
cross sections, we also include the  charm-strange meson pair contribution explicitly (as discussed in the following). More specifically, the $e^+e^-\to D^{(*)+}D^{(*)-}, D^{(*)0}\bar{D}^{(*)0}, D_s^{(*)+}D_s^{(*)-}$ cross sections within the energy region $[3.7, 4.25]~\mathrm{GeV}$ are investigated. Firstly, we present the transformation from the hadronic basis to the SU(3) flavor singlet and octet basis, as well as the isospin triplet basis. Based on the transformation, we can construct the contact potentials with respect to the Heavy Quark Spin Symmetry (HQSS). With these contact potentials, we can solve the Lippmann-Schwinger equation (LSE) to obtain the production amplitudes. In Subsection~\ref{section_C}, we deduce the cross sections formula for the direct comparison with the experimental data. 

\subsection{Transformation between the SU(3)
flavor symmetry basis and $P$-wave hadronic basis}\label{section_A}

Before going into details, we adopt several conventions to facilitate the representation of physical quantities. Unless otherwise specified in the text, $(D^{(*)}\bar{D}^{(*)})^a_n$ denotes charmed meson pairs, with $a=d, u, s$ denoting the light quarks in the charmed meson pairs $(ca)(\bar{c}\bar{a})$ and $n=1,2,3,4$ representing different charmed meson pairs $D\bar{D}$, $D\bar{D}^*$, $D^*\bar{D}^*_{S=0}$ and $D^*\bar{D}^*_{S=2}$. Here, the subindex $S$ of the later two cases is the total spin of the two charmed meson pairs. For instance, $(D^{(*)}\bar{D}^{(*)})^d_1$ denotes the charmed meson pair $D^+D^-$. $(D^{(*)}\bar{D}^{(*)})^a$ denotes all the four charmed meson pairs with light quark pairs $a$. With these conventions, the hadronic basis can be written as $|D^{(*)}\bar{D}^{(*)}\rangle^a_n$. Similarly, the SU(3) flavor basis can be written as $|D^{(*)}\bar{D}^{(*)}\rangle^i_n$, where the index $i=0,8,1$ represents SU(3) singlet $0$, the zero components of octet $8^{00}$ and isospin triplet $1^{10}$, in order,
where the superscripts denote the isospin $I$ and its third component, respectively. Similarly, if the index $n$ is absent, $|D^{(*)}\bar{D}^{(*)}\rangle^i$ denotes all the four charmed meson pairs of the SU(3) representation $i$.

As we consider the cross sections of the charmed meson pairs in electron-positron annihilation, only the third component of various SU(3) flavor representations is produced. As a result, we present the third components of SU(3) singlet and octet as
\begin{align}
|0\rangle=&\frac{1}{\sqrt{3}}\left( |d\bar{d}+u\bar{u}+s\bar{s}\rangle\right),\label{su3_1}\\
|8\rangle =& \frac{1}{\sqrt{6}}\left( |d\bar{d}+u\bar{u}-2s\bar{s}\rangle \right)\label{su3_2},
\end{align}
and the third component of the isospin triplet  
\begin{align}
|1\rangle= \frac{1}{\sqrt{2}}(|d\bar{d}-u\bar{u}\rangle)\label{su3_3}.
\end{align}
The above equations transform into charmed meson pair basis as
\begin{widetext}
    \begin{align}
|D^{(*)}\bar{D}^{(*)}\rangle^0 = & \frac{1}{\sqrt{3}}\left( | D^{(*)+}D^{(*)-}\rangle+|D^{(*)0}\bar{D}^{(*)0}\rangle+|D_s^{(*)+}D_s^{(*)-}\rangle\right),\\
|D^{(*)}\bar{D}^{(*)}\rangle^8 = & \frac{1}{\sqrt{6}}\left( | D^{(*)+}D^{(*)-}\rangle+|D^{(*)0}\bar{D}^{(*)0}\rangle-2|D_s^{(*)+}D_s^{(*)-}\rangle\right),\\
|D^{(*)}\bar{D}^{(*)}\rangle^1 = & \frac{1}{\sqrt{2}}\left( |D^{(*)+}D^{(*)-}\rangle-|D^{(*)0}\bar{D}^{(*)0}\rangle \right).
\end{align}
\end{widetext}
With the above formulae, one can easily transform from the hadronic basis to the SU(3) flavor basis. 

For a given SU(3) representation, one can perform a heavy-light decomposition to obtain the contact potentials. In the HQSS limit, the heavy and light degrees of freedom are conserved individually. The former one is reflected by the total spin $s_Q$ of the heavy quark pair. The latter one is the sum of the total spin $s_q$ of light quark pair and the relative orbital angular momentum $l$ between the two hadrons. Therefore, it is convenient to decompose a charmed meson pair  $|l(\left[s_{l_{1}} s_{Q_{1}}\right]_{j_{1}}\left[s_{l_{2}} s_{Q_{2}}\right]_{j_{2}})_{s}\rangle_{J}$ into the heavy-light basis $|(l\left[s_{l_{1}} s_{l_{2}}\right]_{s_{q}})_{s_{l}}\left[s_{Q_{1}} s_{Q_{2}}\right]_{s_{Q}}\rangle_{J}$, which can be simplified as $|s_Q \otimes s_l\rangle_J$. Here, $s_{l_i}$, $s_{Q_i}$ and $j_i$ are the spin of the light quark plus the relative orbital angular momentum $l$, the heavy quark spin and the total angular momentum of the $i$th mesons, respectively. With this convention, the decomposition read~\cite{Voloshin:2012dk,Du:2016qcr}
\begin{align}   
|l(\left[ s_{l_{1}}\right.& \left. s_{Q_{1}}\right]_{j_{1}}\left[s_{l_{2}} s_{Q_{2}}\right]_{j_{2}})_{s}\rangle_{J} \notag\\
 = & \sum_{s_{l}, s_{Q}, s_{q}}(-1)^{l+s_{q}+s_{Q}+J} \hat{s_{q}} \hat{s_{Q}} \hat{j_{1}} \hat{j_{2}} \hat{s} \hat{s_{l}}\left\{\begin{array}{ccc}
s_{l_{1}} & s_{Q_{1}} & j_{1} \\
s_{l_{2}} & s_{Q_{2}} & j_{2} \\
s_{q} & s_{Q} & S
\end{array}\right\}\notag \\
& \times\left\{\begin{array}{ccc}
l & s_{q} & s_{l} \\
s_{Q} & J & S
\end{array}\right\}|(l\left[s_{l_{1}} s_{l_{2}}\right]_{s_{q}})_{s_{l}}\left[s_{Q_{1}} s_{Q_{2}}\right]_{s_{Q}}\rangle_{J}\notag\\
 = & \sum_{s_{l}, s_{Q}, s_{q}}(-1)^{l+s_{q}+s_{Q}+J} \hat{s_{q}} \hat{s_{Q}} \hat{j_{1}} \hat{j_{2}} \hat{s} \hat{s_{l}}\left\{\begin{array}{ccc}
s_{l_{1}} & s_{Q_{1}} & j_{1} \\
s_{l_{2}} & s_{Q_{2}} & j_{2} \\
s_{q} & s_{Q} & S
\end{array}\right\}\notag \\
& \times\left\{\begin{array}{ccc}
l & s_{q} & s_{l} \\
s_{Q} & J & S
\end{array}\right\}|s_Q \otimes s_l\rangle_J,\label{heavy_light}
\end{align}
with $\hat{j}=\sqrt{2j+1}$. In the Eq.(\ref{heavy_light}), $S$ and $l$ denote the total spin of the two-meson system and its  relative orbital angular momentum, respectively, $J=l+S$ is the total angular momentum. 
As the charmed meson pair $D^{(*)}\bar{D}^{(*)}$ couple to virtual photon, i.e. $J^{PC}=1^{--}$, is in $P$-wave. 
The corresponding decompositions can be obtained from the above equation
\begin{widetext}
    \begin{align}
|D \bar{D}\rangle^i_{1^{--}}  = & p^i \left(D\bar{D}\right)= \frac{1}{2}|0 \otimes 1\rangle^i+\frac{1}{2 \sqrt{3}}|1 \otimes 0\rangle^i-\frac{1}{2}|1 \otimes 1\rangle^i +\frac{1}{2} \sqrt{\frac{5}{3}}|1 \otimes 2\rangle^i,\label{had_1} \\
\left|D \bar{D}^{*}+c . c .\right\rangle^i_{1^{--}} & =\frac{i}{2}\epsilon^{ijk}p_j\left(D_k^{*}\bar{D}-\bar{D}_k^*D\right)= -\frac{1}{\sqrt{3}}|1 \otimes 0\rangle^i+\frac{1}{2}|1 \otimes 1\rangle^i +\frac{1}{2} \sqrt{\frac{5}{3}}|1 \otimes 2\rangle^i,\label{had_2}\\
\left|D^{*} \bar{D}^{*}\right\rangle_{1^{--}}^{i~s= 0} &=\frac{p^i}{\sqrt{3}}\left(D_j^*\bar{D}_j^*\right) = \frac{1}{2} \sqrt{3}|0 \otimes 1\rangle^i-\frac{1}{6}|1 \otimes 0\rangle^i+\frac{1}{2 \sqrt{3}}|1 \otimes 1\rangle^i -\frac{\sqrt{5}}{6}|1 \otimes 2\rangle^i,\label{had_3} \\
\left|D^{*} \bar{D}^{*}\right\rangle_{1--}^{i~s= 2} &=\sqrt{\frac{3}{5}}\frac{p_k}{2}\left(D_i^*\bar{D}_k^*+\bar{D}_i^*D_k^*-\frac{2}{3}\delta_{ik}D_j^*\bar{D}_j^*\right) = \frac{\sqrt{5}}{3}|1 \otimes 0\rangle^i+\frac{1}{2} \sqrt{\frac{5}{3}}|1 \otimes 1\rangle^i +\frac{1}{6}|1 \otimes 2\rangle^i,\label{had_4}
\end{align}
\end{widetext}
which can be represented as a compact transformation  matrix
\begin{align}
    C^{1^{--}}=\begin{pmatrix}\frac{1}{2}&\frac{1}{2\sqrt{3}}&-\frac{1}{2}&\frac{1}{2}\sqrt{\frac{5}{3}}\\\\0&-\frac{1}{\sqrt{3}}&\frac{1}{2}&\frac{1}{2}\sqrt{\frac{5}{3}}\\\\\frac{1}{2}\sqrt{3}&-\frac{1}{6}&\frac{1}{2\sqrt{3}}&-\frac{\sqrt{5}}{6}\\\\0&\frac{\sqrt{5}}{3}&\frac{1}{2}\sqrt{\frac{5}{3}}&\frac{1}{6}\end{pmatrix}.
\end{align}
The wave functions on the left side of Eqs.~\eqref{had_1}-\eqref{had_4}, are the hadronic basis with $\vec{p}$ the three momentum of final particle in the center-of-mass frame. The wave functions on the right side are in the heavy-light basis.
One can easily check that these bases are normalized and orthogonal to each other from both the heavy-light basis and the hadronic basis~\cite{Voloshin:2012dk,Du:2016qcr} in Eqs.~\eqref{had_1}-\eqref{had_4}. In the hadron basis, one can see that the wave functions are normalized to $|\vec{p}|^2$. The positive sign in $|D\bar{D}^*+c.c.\rangle^i_{1^{--}}$ is related to the $C$-parity, where we adopt the convention $D\stackrel{\mathcal{C}}\longrightarrow \bar{D}$, $D^*\stackrel{\mathcal{C}}\longrightarrow -\bar{D}^*$. The transformation between hadronic basis $|D^{(*)}\bar{D}^{(*)}\rangle^a$ and SU(3) flavor symmetry basis $|D^{(*)}\bar{D}^{(*)}\rangle^i$ is
\begin{align}
&[|D^{(*)+}D^{(*)-}\rangle^d, |D^{(*)0}\bar{D}^{(*)0}\rangle^u, |D_s^{(*)+}D_s^{(*)-}\rangle^s]^T\notag\\
&=R[|D^{(*)}\bar{D}^{(*)}\rangle^0,|D^{(*)}\bar{D}^{(*)}\rangle^8,|D^{(*)}\bar{D}^{(*)}\rangle^1]^T,
\end{align}
where the transformation matrix $R$ is
\begin{align} 
    R=\left(\begin{array}{ccc}
    \frac{1}{\sqrt{3}} &\frac{1}{\sqrt{6}} &\frac{1}{\sqrt{2}} \\
     \frac{1}{\sqrt{3}} &\frac{1}{\sqrt{6}} &\frac{-1}{\sqrt{2}}\\
      \frac{1}{\sqrt{3}}&\frac{-2}{\sqrt{6}}&0
    \end{array}\right)\otimes \bm{1}_{4\times 4},\label{r21}
\end{align}
with $\bm{1}_{4\times 4}$ the $4\times 4$ identity matrix.

In the HQSS limit, one can define the low-energy constants
\begin{align}
C_{1}^i &\equiv  {^i\langle} 0 \otimes 1|\mathcal{H}_{CT}| 0 \otimes 1\rangle^j\delta_{ij},\label{con_1}\\
C_{2}^i &\equiv  {^i\langle} 1 \otimes 0|\mathcal{H}_{CT}| 1 \otimes 0\rangle^j\delta_{ij},\label{con_2}\\
C_{3}^i &\equiv  {^i\langle} 1 \otimes 1|\mathcal{H}_{CT}| 1 \otimes 1\rangle^j\delta_{ij},\label{con_3}\\
C_{4}^i& \equiv  {^i\langle} 1 \otimes 2|\mathcal{H}_{CT}| 1 \otimes 2\rangle^j\delta_{ij},\label{con_4}
\end{align}
where the repeated indices do not imply summation. Here, $\mathcal{H}_{CT}$ represents the leading order Hamiltonian which respects HQSS. Since we focus on the energy region $[3.7, 4.25]~\mathrm{GeV}$, only the leading order contact potentials 
$C_{1,2,3,4}^i$ are considered as constants within such a small energy region. The contact potentials read 
\begin{align}
    V_{nn'}^i= {^i_n\langle}D^{(*)}\bar{D}^{(*)}|\mathcal{H}_{CT}|D^{(*)}\bar{D}^{(*)}\rangle^j_{n'}\delta_{ij}.\label{CT}
\end{align}
Substituting Eq.~(\ref{had_1})-Eq.~(\ref{had_4}) into Eq.~(\ref{CT}) and combining with Eq.~(\ref{con_1})-Eq.~(\ref{con_4}), one can obtain the explicit form of $V_{nn'}^i$
\begin{widetext}
\begin{align}
V_{11}^i & = \frac{1}{4}C^i_1+\frac{1}{12} C_{2}^i+\frac{1}{4} C_{3}^i+\frac{5}{12} C_{4}^i ,& V_{12}^i  &= -\frac{1}{6} C_{2}^i-\frac{1}{4} C_{3}^i+\frac{5}{12} C_{4}^i, \notag\\
V_{13}^i & = \frac{\sqrt{3}}{4}C_1^i-\frac{1}{12 \sqrt{3}} C_{2}^i+\frac{\sqrt{3}}{12} C_{3}^i-\frac{5 \sqrt{3}}{36} C_{4}^i,& V_{14}^i & = \frac{1}{6} \sqrt{\frac{5}{3}} C_{2}^i-\frac{1}{4} \sqrt{\frac{5}{3}} C_{3}^i+\frac{1}{12} \sqrt{\frac{5}{3}} C_{4}^i,\notag \\
V_{22}^i & = \frac{1}{3} C_{2}^i+\frac{1}{4} C_{3}^i+\frac{5}{12} C_{4}^i ,&  V_{23}^i & = \frac{1}{6 \sqrt{3}} C_{2}^i+\frac{1}{4 \sqrt{3}} C_{3}^i-\frac{5}{12 \sqrt{3}} C_{4}^i,\notag \\
V_{24}^i & = -\frac{1}{3} \sqrt{\frac{5}{3}} C_{2}^i+\frac{1}{4} \sqrt{\frac{5}{3}} C_{3}^i+\frac{1}{12} \sqrt{\frac{5}{3}} C_{4}^i ,& V_{33}^i & =\frac{3}{4}C_1^i+ \frac{1}{36}C_2^i+\frac{1}{12} C_{3}^i+\frac{5}{36} C_{4}^i, \notag\\
V_{34}^i & = -\frac{\sqrt{5}}{18} C_{2}^i+\frac{\sqrt{5}}{12} C_{3}^i-\frac{\sqrt{5}}{36} C_{4}^i ,& V_{44}^i & = \frac{5}{9} C_{2}^i+\frac{5}{12} C_{3}^i+\frac{1}{36} C_{4}^i.\label{contact_potential}
\end{align}
\end{widetext}
Since $V^i$ is a symmetric $4\times 4$ matrix, we only show the elements $V_{nn'}$ with $n < n'$. One might have noticed that the $D^{(*)}\bar{D}^{(*)}$ pairs are formed to a $J^{PC}=1^{--}$ state in $P$-wave, which should encode a momentum dependence in each vertex, reflecting the $P$-wave interaction. Here, we use a separable contact interaction. The momentum in the loop is contained in the two-point propagator, and the momentum dependence of the external particles is contained in the amplitude, which will be discussed afterwards.
The contact potentials in the SU(3) flavor basis reads
\begin{align}
    V_{\mathrm{CT}}=\left(\begin{array}{ccc}
        V^0 &  &  \\
         & V^8 &  \\
         &  & V^1 \\
    \end{array}\right),
\end{align}
where $V^0$, $V^8$ and $V^1$ are $4\times 4$ matrices whose elements are given by the Eq.~(\ref{contact_potential}).

Within the energy region of interest $[3.7,4.25]~\mathrm{GeV}$, there are three vector charmonia, i.e. $\psi(1D)$, $\psi(3S)$, $\psi(2D)$. The vector charmonium $\psi(2S)$ is slightly below the $D\bar{D}$ threshold and might also affect the cross sections of two charmed meson pairs. However, we have checked that its effect is marginal and we neglect its contribution in our framework. In this case, the vector charmonia $\psi(1D)$, $\psi(3S)$ and $\psi(2D)$ are included one by one.  
As a result, we consider three different frameworks, 
i.e. the coupled-channel effect with the three vector charmonia (Model I), two vector charmonia (Model II), and 
one vector charmonium (Model III).
From the experimental side, one can see 
significant contributions of the $\psi(1D)$ and $\psi(3S)$ in 
the $e^+e^-\rightarrow D^+D^-/D^0\bar{D}^0$~\cite{julin2017measurement,Belle:2007qxm} and $e^+e^-\rightarrow D^+\bar{D}^{*-}$ \cite{Belle:2017grj,BESIII:2021yvc} processes, respectively. Therefore, we need to consider  Model I and Model II, without focusing on Model III. The $S$-wave and $D$-wave charmonia can be expressed in terms of heavy-light basis $|1\otimes 0\rangle$ and $|1\otimes 2\rangle$, respectively. 
As these charmonia are SU(3) flavor singlet, we define the coupling constants between charmonia and the open charmed channels 
\begin{align}
g_{1D}^0 &\equiv  {^0\langle} 1 \otimes 2|\mathcal{H}_{\text{bare}}| 1 \otimes 2\rangle^0_{1D},\\
g_{3S}^0 &\equiv {^0\langle} 1 \otimes 0|\mathcal{H}_{\text{bare}}| 1 \otimes 0\rangle^0_{3S},\\
g_{2D}^0 &\equiv {^0\langle} 1 \otimes 2|\mathcal{H}_{\text{bare}}| 1 \otimes 2\rangle^0_{2D}.
\end{align}
 The heavy-light structures  $|1\times 2\rangle_{1D}^0$, $|1\times 0\rangle_{3S}^0$ and $|1\times 2\rangle_{2D}^0$ corresponds to $\psi(1D)$, $\psi(3S)$ and $\psi(2D)$, respectively. $\mathcal{H}_{\text{bare}}$ is the Hamiltonian density describing the interaction between the bare state and the open charmed meson pair, with the corresponding potential 
\begin{align}
    V_{c\bar{c}~n j}^0= {^0_n\langle} D^{(*)}\bar{D}^{(*)}|\mathcal{H}_{\text{bare}}|j{\rangle^0},
\end{align}
where $j=1,2,3$ ($j=1,2$) denote charmonia $\psi(1D)$, $\psi(3S)$, $\psi(2D)$ for Model I ($\psi(1D)$, $\psi(3S)$ for Model II).

\subsection{The Lippmann-Schwinger equation}
In total, there are $12+\alpha$ channels in our framework, with $12$ open charmed channels and 
$\alpha$ bare vector charmonium states ($\alpha=3$ or 2, depending on the Moldel I or II). The Lippmann-Schwinger equation (LSE) reads as 
\begin{align}
    T(E) = V + VG(E)T(E),\label{LSE}
\end{align}
where $E$ is the total energy in the center-of-mass (c.m.) frame.
$V$ and $G(E)$ denote the potential and two-point loop function matrices.
The potential $V$ reads as 
\begin{align}
    V = \left(\begin{array}{cc}
\left[V_{\text{oo}}\right]_{12 \times 12} & \left[V_{\text{ob}}\right]_{12 \times {\alpha}} \\
\left[V_{\text{bo}}\right]_{{\alpha} \times 12} & \bm{0}_{{\alpha}\times {\alpha}}
\end{array}\right),\label{potential_V}
\end{align}
where $V_{\text{oo}}$ represents the contact potential between two charmed meson channels, and $V_{\text{ob}}$ denotes the interaction between the bare vector charmonium and the charmed meson pairs. The two-point function matrix $G(E)$ reads~\cite{Guo:2015umn}
\begin{align}
    G(E) = \left( \begin{array}{cc}
       \text{diag}\left[G_{CT}^{ii}(E)\right]_{12 \times 12} & \bm{0}_{12\times {\alpha}} \\
       \bm{0}_{{\alpha}\times 12} & \text{diag}\left[G_{c\bar{c}}(E)\right]_{{\alpha}\times {\alpha}} \\
    \end{array} \right),\label{two-body_fun}
\end{align}
with
\begin{align}
& G^{ii}_{CT}(E) = \int \frac{d^{3} \vec{q}}{(2 \pi)^{3}} \frac{q^{2} f^2_{\Lambda}\left(q^{2}\right)}{E-m_{i1}-m_{i2}-q^{2} /(2 \mu)+i\varepsilon^+}\notag \\
& = -\frac{\mu \Lambda}{(2 \pi)^{3 / 2}}\left(k^{2}+\frac{\Lambda^{2}}{4}\right)+\frac{\mu k^{3}}{2 \pi} e^{-2 k^{2} / \Lambda^{2}}\left[\operatorname{erfi}\left(\frac{\sqrt{2} k}{\Lambda}\right)-i\right],\notag\\
&G_{c\bar{c}}(E) =\frac{1}{E^2-m^2+i\varepsilon^+},
\end{align}
where $m$ and $\Lambda$ denote the charmonium bare mass and cutoff, respectively. $m_{i1}$ and $m_{i2}$ are the meson masses involved in the $i$th channel. Here, we take the Gaussian form factor $f_{\Lambda}(q^2)=\mathrm{exp}(-q^2/{\Lambda^2})$. The momentum $q^2$ in the numerator reflects the $P$-wave interaction between the charmed meson pairs. More details of the $P$-wave two-point function $G_{CT}^{ii}(E)$ can be found in App.~\ref{Green_fun}. We employ the nonrelativistic Green function because the relevant dynamics occur near thresholds. The relativistic correction can be estimated by $p^2/4(2\mu)^2$ with $p$ and $\mu$ the c.m. three momentum and reduced mass of a given channel. This estimate is from the expansion of the energy $E=m+\frac{p^2}{2m}+\frac{1}{8}m\frac{p^4}{m^4}+\cdots$ in terms of momentum and mass, where ratio between the third term and the second term is $p^2/4m^2$. For the two-body channel, we replace the mass $m$ by $2\mu$. 
 This value for the lowest channel at the highest energy is about $0.06$ at the amplitude level, which means that the relativistic correction is at most $1.06^2-1=12\%$ for physical quantities. On the other hand, both the relativistic and non-relativistic expressions should be compared with the experimental data. At the end, part of this correction will be absorbed into the redefinition of the model parameters. From this point of view,  $12\%$ is the maximum estimate of the relativistic correction. This indicates that the nonrelativistic approximation remains valid for the majority of channels throughout the energy region of interest. Although relativistic effects may become significant in energy regions far from the thresholds, they do not affect the physical results near the thresholds. 
Substituting Eq.~(\ref{potential_V}) and Eq.~(\ref{two-body_fun}) into Eq.~(\ref{LSE}), one can obtain
\begin{widetext}
\begin{align}
\left(\begin{array}{cc}
\left[T_{\text{oo}} \right]_{12 \times 12}& \left[T_{\text{ob}} \right]_{12 \times {\alpha}}\\
\left[T_{\text{bo}} \right]_{{\alpha}\times 12}& \left[T_{\text{bb}}\right]_{{\alpha}\times {\alpha}}
\end{array}\right) & = \left(\begin{array}{cc}
\left[V_{\text{oo}}+V_{\text{oo}} G_{CT} T_{\text{oo}}+V_{\text{ob}} G_{c\bar{c}} T_{\text{bo}}\right]_{12\times 12} & \left[V_{\text{ob}}+V_{\text{oo}} G_{CT} T_{\text{ob}}+V_{\text{ob}} G_{c\bar{c}} T_{\text{bb}}\right]_{12\times {\alpha}} \\
\left[V_{\text{bo}}+V_{\text{bo}} G_{CT} T_{\text{oo}}\right]_{{\alpha}\times 12} & \left[V_{\text{bo}} G_{CT} T_{\text{ob}}\right]_{{\alpha}\times {\alpha}}
\end{array}\right),
\end{align}
\end{widetext}
where $T_{\text{oo}}$ denotes the scattering amplitudes between charmed meson pairs. $T_{\text{ob}}$ denotes the scattering amplitudes between the charmonium and the charmed meson pair, with $T_{\text{bo}}$ describes the inverse process of $T_{\text{ob}}$, and $G_{CT}=\text{diag}\left[G_{CT}^{ii}(E)\right]_{12\times 12}$. Plugging the $T_{\text{bo}}$ into $T_{\text{oo}}$, one  obtains
\begin{align}
    T_{\text{oo}}(E) = \hat{V}_{\text{oo}}^{\text{eff}}+\hat{V}_{\text{oo}}^{\text{eff}}G_{CT}(E)T_{\text{oo}}(E),
\end{align}
where the effective potential is defined as $\hat{V}_{\text{oo}}^{\text{eff}}\equiv V_{\text{oo}}+V_{\text{ob}}G_{c\bar{c}}V_{\text{bo}}$. It is easy to see that
\begin{align}
    T_{\text{oo}}(E) = \left[\left[\hat{V}_{\text{oo}}^{\text{eff}}(E)\right]^{-1} - G_{CT}(E) \right]^{-1}\label{T-martix}
\end{align}
by solving algebraic LSE. 
To ensure  unitarity of the $T$-matrix, the Gaussian form factor $f_{\Lambda}(p)$ appearing in the two-point loop function $G^{ii}_{CT}(E)$ should also contribute to the above $T$-matrix as~\footnote{This form factor can also be added to each vertex in the potential alternatively, instead in the two-point loop and external particles, to satisfy the unitarity is automatically. } 
\begin{align}
    T_{\text{oo}}(E) = f_{\Lambda}(p)\left[\left[\hat{V}_{\text{oo}}^{\text{eff}}(E)\right]^{-1} - G_{CT}(E) \right]^{-1}f_{\Lambda}(p').\label{T_matrix}
\end{align}

Since these charmonia only couple to the SU(3) flavor singlet, the effective potential can be represented as
\begin{align}
\hat{V}_{\text{oo}}^{\text{eff}'}=\left(\begin{array}{ccc}
        V^0+V_{c\bar{c}}^0 G_{c\bar{c}} {V_{c\bar{c}}^0}^T &  &  \\
         & V^8 &  \\
         &  & V^1 \\
    \end{array}\right).\label{V-eff}    
\end{align}
In contrast to the effective potential (Eq.~(\ref{T_matrix})) in particle basis, Eq.~(\ref{V-eff}) is expressed in the SU(3) flavor basis. One needs to transform $\hat{V}_{\text{oo}}^{\text{eff}'}$ into particle basis
\begin{align}
    \hat{V}_{\text{oo}}^{\text{eff}}= R \hat{V}_{\text{oo}}^{\text{eff}'} R^{-1},\label{V_eff}
\end{align}
with the transformation matrix $R$ given in Eq.~(\ref{r21}).
Substituting Eq.~(\ref{V_eff}) into Eq.~(\ref{T_matrix}), one can obtain the full mesonic $T$-matrix for the coupled-channel system $(D^{(*)}\bar{D}^{(*)})^a$ containing the contributions from bare charmonium states.

\subsection{The physical production amplitude and cross section}\label{section_C}

Analogous to the LSE, the physical production amplitude reads
\begin{align}
    \mathcal{U}(E)=\mathcal{F}+VG(E)\mathcal{U}(E),\label{PA}
\end{align}
where $\mathcal{F}=(\left[F_\text{o}\right]_{12\times 1}^T,\left[f_\text{b}\right]^T_{n\times 1})^T$ is the bare production amplitude. The $F_{\text{o}}$ matrix is the bare production between the virtual photon and charmed meson pair. The $f_{\text{b}}$ matrix is the bare production between the virtual photon and the charmonia. Similarly, plugging Eq.~(\ref{potential_V}) and Eq.~(\ref{two-body_fun}) into Eq.~(\ref{PA}), one obtains the physical production explicitly  
\begin{align}
    \left(\begin{array}{c}
         \left[\mathcal{U}_{\text{o}}\right]_{12\times 1}  \\
          \left[\mathcal{U}_{\text{b}}\right]_{\alpha\times 1}
    \end{array}\right)=\left(\begin{array}{c}
\left[F_{\text{o}}+V_{\text{oo}}G_{CT}\mathcal{U}_{\text{o}}+V_{\text{ob}}G_{c\bar{c}}\mathcal{U}_{\text{b}}\right]_{12\times 1}  \\
\left[f_{\text{b}}+V_{\text{bo}}G_{CT}\mathcal{U}_{\text{o}}\right]_{\alpha\times 1}
    \end{array}\right),
\end{align}
where $\mathcal{U}_{\text{o}}$ and $\mathcal{U}_{\text{b}}$ represent the physical production amplitudes for the charmed meson pairs and the involved charmonia, respectively. Substituting $\mathcal{U}_{\text{b}}$ into $\mathcal{U}_{\text{o}}$, one we get the physical production amplitude for the open charmed channels
\begin{align}
    \mathcal{U}_{\text{o}}(E) = (\bm{1}_{12\times 12}-\hat{V}_{\text{oo}}^{\text{eff}}G_{CT}(E))^{-1}\hat{F}_{\text{o}}^{\text{eff}},\label{P_A}
\end{align}
with $\hat{F}_{\text{o}}^{\text{eff}}\equiv F_{\text{o}}+V_{\text{ob}}G_{c\bar{c}}f_{\text{b}}$
the effective bare production amplitude. Analogous to that for $T$-matrix, the Gaussian form factor $f_{\Lambda}(p)$ is introduced to regularize the integral. As the result, Eq.~(\ref{P_A}) can be rewritten as
\begin{align}
    \mathcal{U}_{\text{o}}(E) = f_{\Lambda}(p)(\bm{1}_{12\times 12}-\hat{V}_{\text{oo}}^{\text{eff}}G_{CT}(E))^{-1}\hat{F}_{\text{o}}^{\text{eff}}.\label{P_A}
\end{align}
As the standard QED vertex between the virtual photon and the $c\bar{c}$ state can be decomposed into both $S$-wave and $D$-wave $c\bar{c}$ pairs~\cite{Li:2013yka,Wang:2013kra}, one can define the coupling between the virtual photon and the $m$-th open charmed channel in the same SU(3) flavor basis as
\begin{align}
    F^i_m \equiv C^{1^{--}}_{m2}f_S^i+C^{1^{--}}_{m4}f_D^i,
\end{align}
with $m=1,2,3,4$, and $
    f_S^i\equiv {^i\langle}1\otimes 0|\mathcal{H}_{\text{EM}}|\gamma^*\rangle^i,
    f_D^i\equiv {^i\langle}1\otimes 2|\mathcal{H}_{\text{EM}}|\gamma^*\rangle^i.$ Here, $\mathcal{H}_{\text{EM}}$ is the electromagnetic Hamiltonian density describing the interaction between the virtual photon and vector charmonia. The explicit form of the bare production (for Model I) amplitude reads
\begin{align}
    \mathcal{F} & = (\left[F_{\text{o}}\right]^T_{12\times 1},\left[f_{\text{b}}\right]^T_{3\times 1})^T\notag\\
    & =(\left[F^0\right]_{4\times 1}^T,\left[F^8\right]_{4\times 1}^T,\left[F^1\right]_{4 \times 1}^T,\left[f_{\text{b}}\right]_{3\times 1}^T)^T,
\end{align}
where $F^i=(F^i_1,F^i_2,F^i_3,F^i_4)^T$ and $f_{\text{b}}=(f_{1D}^0,f_{3S}^0,f_{2D}^0)^T$ for Model I ($f_{\text{b}}=(f_{1D}^0,f_{3S}^0)^T$ for Model II) which is the couplings between the virtual photon and charmonia. The effective bare production amplitude is given by
\begin{align}
    \hat{F}_{\text{o}}^{\text{eff}'} = \left( \begin{array}{c}
       \left[F^0+V_{\text{ob}}G_{c\bar{c}}f_{\text{b}} \right]_{4\times 1}   \\
        \left[F^8\right]_{4\times 1} \\
        \left[F^1\right]_{4\times 1}
    \end{array}\right).\label{F-eff}
\end{align}
Similarly, we need to transform $\hat{F}_{\text{o}}^{\text{eff}'}$ in Eq.(\ref{F-eff}) into the particle basis representation
\begin{align}
    \hat{F}_{\text{o}}^{\text{eff}} = R\hat{F}_{\text{o}}^{\text{eff}'}.\label{F_eff}
\end{align}
Substituting Eq.~(\ref{V_eff}) and Eq.~(\ref{F_eff}) into Eq.~(\ref{P_A}), one can obtain the physical production amplitude for the open charmed channel containing the three charmonia.
\begin{figure}
    \centering    \includegraphics[width=0.9\linewidth]{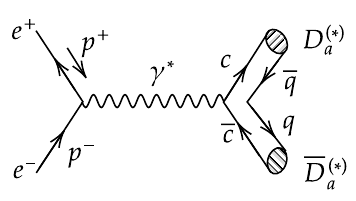}
    \caption{Feynman diagram for the processes $e^+e^-\rightarrow (D^{(*)}\bar{D}^{(*)})_a$.}   
    \label{FD}
\end{figure}
The process $e^+e^-\rightarrow (D^{(*)}\bar{D}^{(*)})^a_n$ is shown in Fig.~\ref{FD}, and the scattering amplitude for the $n$-th process is given by
\begin{align}
    \mathcal{M}_n^a & = \bar{v}(p_+)(-ie\gamma_\mu)u(p_-)\frac{-ig^{\mu\nu}}{s+i\varepsilon^+}\mathcal{A}_{n\nu}^a,
    \label{sca_amp}
\end{align}
where $p_+(p_-)$ is the four-momentum of the positron (electron) and $s$ is the square of the center-of-mass energy. $\mathcal{A}_{n\nu}^{a}$ is the physical production amplitude with $\nu$ the polarization index. Then the square of the scattering amplitude is
\begin{widetext}
\begin{align}
    |\overline{\mathcal{M}_n^a}|^2 =\frac{1}{2}\sum_r\frac{1}{2}\sum_s\sum_{\lambda}\sum_{\lambda'}|\mathcal{M}_n^a|^2
    & = \frac{e^2}{4s^2}\sum_r\sum_s\sum_{\lambda}\sum_{\lambda'}\bar{v}^r(p_+)\gamma^{\nu}u^s(p_-)\bar{u}^s(p_-)\gamma^{\nu'}v^r(p_+)A^a_{n\nu}A^{a*}_{n\nu'}\notag\\
    & = 
    -\frac{4\pi\alpha}{s^2}\sum_{\lambda}\sum_{\lambda'}(\frac{1}{2}sg^{ij}+2p_+^ip_+^j)A^a_{ni}A_{nj}^{a*} \quad\quad (i,j = 1,2,3),\label{Amp}
\end{align}
\end{widetext}
where we average over the initial electron and positron spins and sum over the polarization of the final charmed meson. Then $\mathcal{A}_n^{ai}$ is given by \cite{Chung:1993da,Zou:2002ar, Voloshin:2012dk} (also see App.~\ref{pro_amp})
\begin{align}
        \mathcal{A}_1^{ai}&=\mathcal{U}_1^a(p_{\bar{D}_a}-p_{D_a})^i, \label{Amp_1}\\
        \mathcal{A}_2^{ai}&=\mathcal{U}_2^a\epsilon^{ijk}(p_{\bar{D}_a}-p_{D^*_a})_j\varepsilon_{\lambda k}^*,\label{Amp_2} \\
        \mathcal{A}_3^{ai}&=\frac{1}{\sqrt{3}}\mathcal{U}_3^a(p_{\bar{D}^*_a}-p_{D^*_a})^i\varepsilon^{*\alpha}_{\lambda}\varepsilon^*_{\lambda' \alpha},\label{Amp_3}\\
        \mathcal{A}_4^{ai}&=\sqrt{\frac{3}{5}}\mathcal{U}_4^aP^{ij,mn}_2(p_{\bar{D}^*_a}-p_{D^*_a})_j\varepsilon^*_{\lambda m}\varepsilon^*_{\lambda' n},\label{Amp_4}
\end{align}
with $P_2^{ij,mn}=\frac{1}{2}\delta^{im}\delta^{jn}+\frac{1}{2}\delta^{in}\delta^{jm}-\frac{1}{3}\delta^{ij}\delta^{mn}$. Here, $p_{D^{(*)}_a}$ and $\varepsilon^*$ are the four-momentum and the polarization vector of the charmed meson, respectively. The momentum dependence in the above four equations reflects the $P$-wave interaction of the charmed meson pairs. More detail can be found in App.~\ref{pro_amp}. Substituting Eqs.~(\ref{Amp_1})-(\ref{Amp_4}) into Eq.~(\ref{Amp}), one can obtain the explicit form of the corresponding amplitudes squared
\begin{align}
    |\overline{\mathcal{M}_1^a}|^2 
    & = \frac{8\pi\alpha}{s}|p_{D^a}|^2|\mathcal{U}_1^a|^2(1-\mathrm{cos}^2\theta),\label{M_1}\\
    |\overline{\mathcal{M}_2^a}|^2 
    & = \frac{8\pi\alpha}{s}|p_{D^a}|^2|\mathcal{U}_2^a|^2(1+\mathrm{cos}^2\theta),\label{M_2}\\
    |\overline{\mathcal{M}_3^a}|^2  
    & = \frac{8\pi\alpha}{s}|p_{D^{*a}}|^2|\mathcal{U}_3^a|^2(1-\mathrm{cos}^2\theta),\label{M_3}\\
    |\overline{\mathcal{M}_4^a}|^2 
    & = \frac{28\pi\alpha}{5s}|p_{D^{*a}}|^2|\mathcal{U}_4^a|^2(1-\frac{1}{7}\mathrm{cos}^2\theta),\label{M_4}
\end{align}
with the fine-structure constant $\alpha=\frac{e^2}{4\pi}$. Here, $\theta$ is the relative angle between the incoming electron and outgoing charmed meson. More details can be found in Appendix B. For the two-body scattering, the differential cross section is given by
\begin{align}
    \frac{d\sigma^a_n}{d\mathrm{cos}\theta}=\frac{|p_{D^{(*)a}}|}{16\pi s^{3/2}}|\overline{\mathcal{M}^a_n}|^2.\label{dsigma}
\end{align}
One can obtain the total cross section 
\begin{align}
    \sigma_1^a=\frac{2\alpha|p_{D^a}|^3}{3s^{5/2}}|\mathcal{U}_1^a|^2,\\
    \sigma_2^a=\frac{4\alpha|p_{D^a}|^3}{3s^{5/2}}|\mathcal{U}_2^a|^2,\\
    \sigma_3^a=\frac{2\alpha|p_{D^{*a}}|^3}{3s^{5/2}}|\mathcal{U}_3^a|^2,\\
    \sigma_4^a=\frac{2\alpha|p_{D^{*a}}|^3}{3s^{5/2}}|\mathcal{U}_4^a|^2,
\end{align}
by plugging Eqs.~(\ref{M_1})-(\ref{M_4}) into Eq.~(\ref{dsigma}) and integrating over the angle $\theta$. 

\subsection{The $1^{-+}$ $P$-wave coupled channel system}

The scattering between two heavy-quark-spin multiplets can be described by the same set of low-energy constants, 
which relates the dynamics of various systems to each other. 
For the $P$-wave scattering between the $s_l^P=\frac{1}{2}^-$ doublet, i.e. $(D, D^*)$ doublet, and its anti-doublet, there is also $J^{PC}=1^{-+}$ exotic quantum number~\cite{Du:2016qcr,Zhang:2025gmm} in addition to the vector channel, i.e. $J^{PC}=1^{--}$.  
For this exotic quantum number, which is beyond the conventional quark model, there is no conventional charmonium coupling to this channel. The dynamic of this quantum number is described by the low-energy constants. Similarly to that in vector channel, the $1^{-+}$ hadronic basis can be represented by the heavy-light basis as 
\begin{align}
    | D\bar{D}^*+c.c.\rangle_{1^{-+}}^i & =-\frac{1}{\sqrt{2}}|0\otimes 1\rangle^i+\frac{1}{\sqrt{2}}|1\otimes 1\rangle^i,\\
    | D^*\bar{D}^*\rangle_{1^{-+}}^{i~s=1} & =\frac{1}{\sqrt{2}}|0\otimes 1\rangle^i+\frac{1}{\sqrt{2}}|1\otimes 1\rangle^i,
\end{align}
where only $|0 \otimes 1\rangle^i$ and $|1\otimes 1\rangle^i$ components appear. The contact potential reads
\begin{align}
    V^{1^{-+}}=\left(\begin{array}{ccc}
        V^0_{1^{-+}} &  &  \\
         & V^8_{1^{-+}} &  \\
         &  & V^1_{1^{-+}} \\
    \end{array}\right),
\end{align}
where $V^i_{1^{-+}}$ is given by
\begin{align}
    V^i_{1^{-+}} = \left(\begin{array}{cc}
       \frac{C_1^i}{2}+\frac{C_3^i}{2}  & -\frac{C_1^i}{2}+\frac{C_3^i}{2} \\
        -\frac{C_1^i}{2}+\frac{C_3^i}{2} & \frac{C_1^i}{2}+\frac{C_3^i}{2}
    \end{array}\right),
\end{align}
which is represented in the SU(3) flavor basis. Similarly, we need to transform the above contact potential into the particle basis
\begin{align}
    V_{CT}^{1^{-+}}=R' \left(\begin{array}{cc}
       \frac{C_1^i}{2}+\frac{C_3^i}{2}  & -\frac{C_1^i}{2}+\frac{C_3^i}{2} \\
        -\frac{C_1^i}{2}+\frac{C_3^i}{2} & \frac{C_1^i}{2}+\frac{C_3^i}{2}
    \end{array}\right)R'^{-1},
\end{align}
with 
\begin{align}
     R^\prime=\left(\begin{array}{ccc}
    \frac{1}{\sqrt{3}}&\frac{1}{\sqrt{6}}&\frac{1}{\sqrt{2}}\\
     \frac{1}{\sqrt{3}}&\frac{1}{\sqrt{6}}&\frac{-1}{\sqrt{2}}\\
      \frac{1}{\sqrt{3}}&\frac{-2}{\sqrt{6}}&0
    \end{array}\right) \otimes \bm{1}_{2\times 2}.
\end{align}
The $T$-matrix of $1^{-+}$ system is given by
\begin{align}
    T(E)_{1^{-+}}=f_{\Lambda}(p)\left[\left[V_{CT}^{1^{-+}}\right]^{-1}-G_{CT}(E)\right]^{-1}f_{\Lambda}(p').\label{T_1np}
\end{align}
One can also study the pole structure of the $1^{-+}$ system through the $T$-matrix, which involves solving $\textbf{det}\left[\bm{1}_{6 \times 6}-V_{CT}^{1^{-+}}G_{CT}(E)\right]=0$.

\section{RESULTS AND DISCUSSION}
\label{sec:results}

\begin{figure*}
    \centering
    \includegraphics[width=0.432\linewidth]{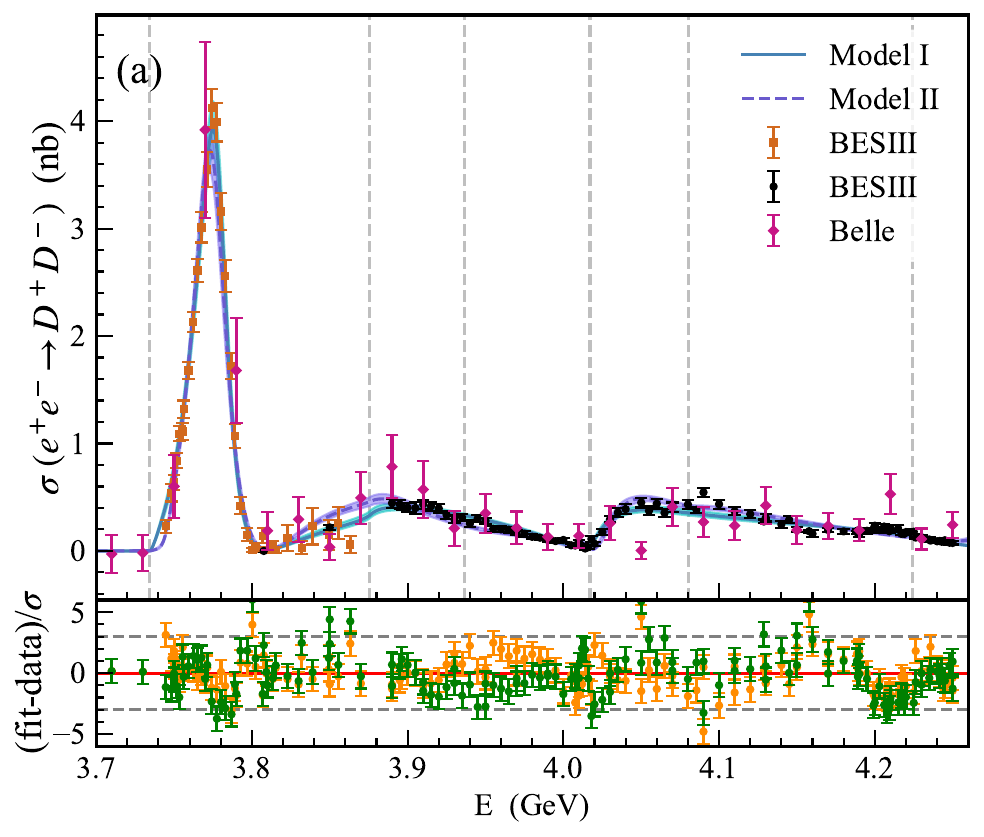}
    \includegraphics[width=0.432\linewidth]{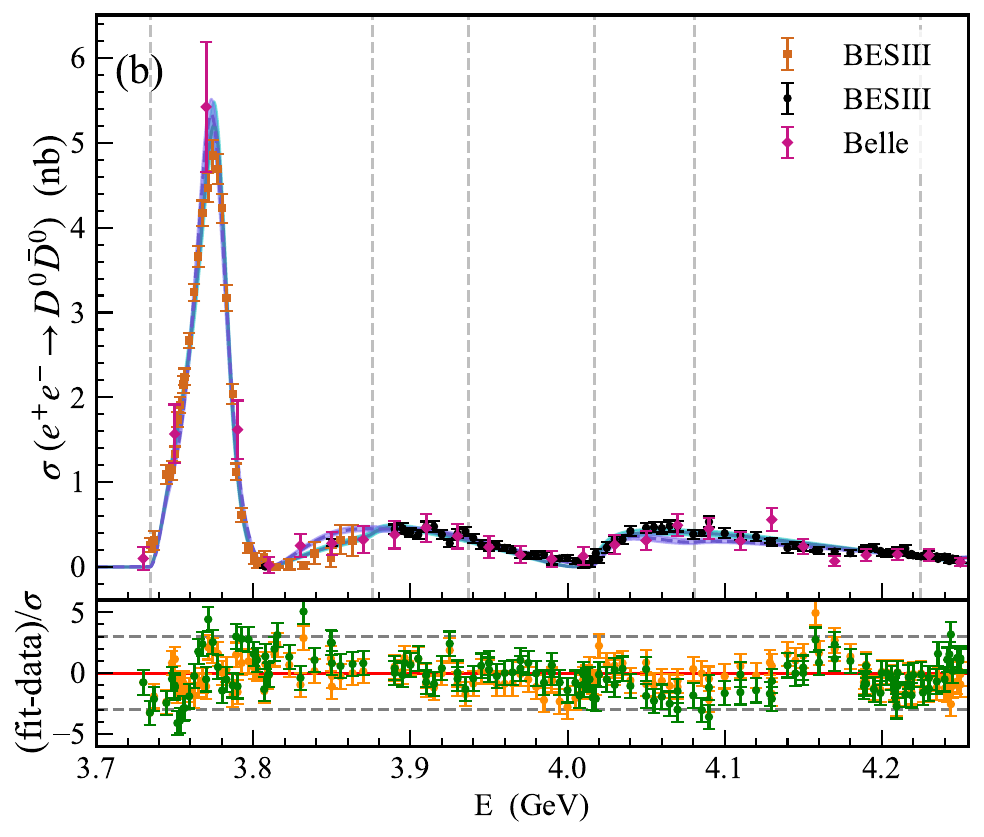}
    \includegraphics[width=0.432\linewidth]{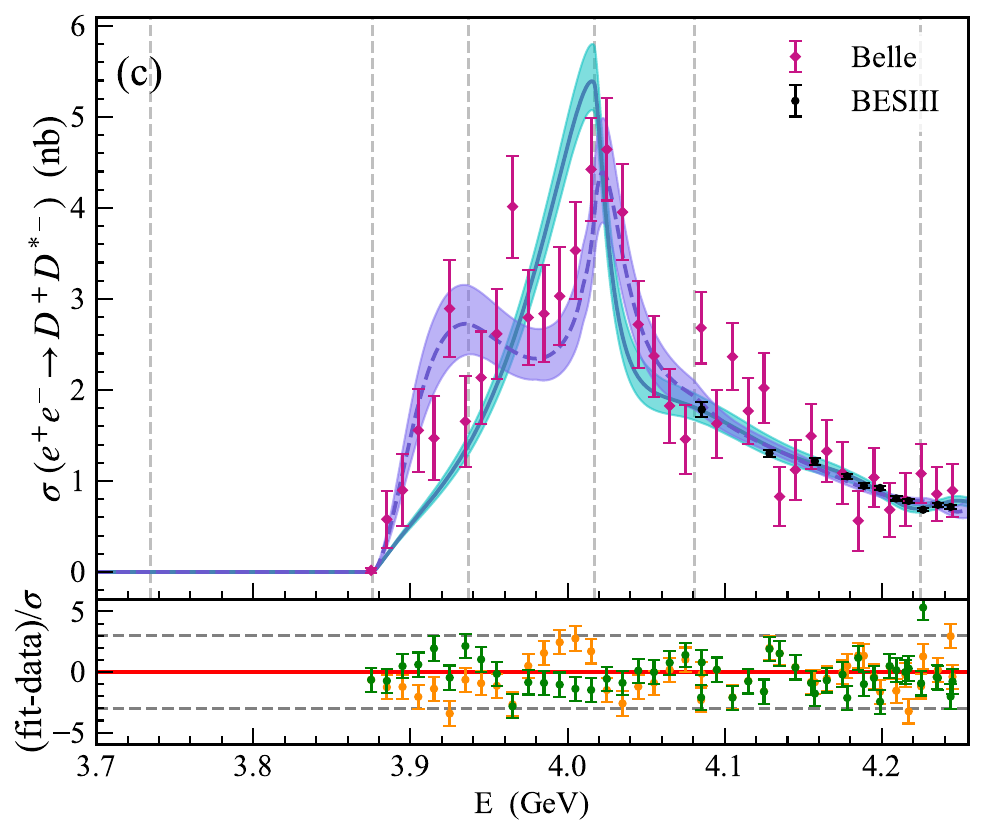}
    \includegraphics[width=0.432\linewidth]{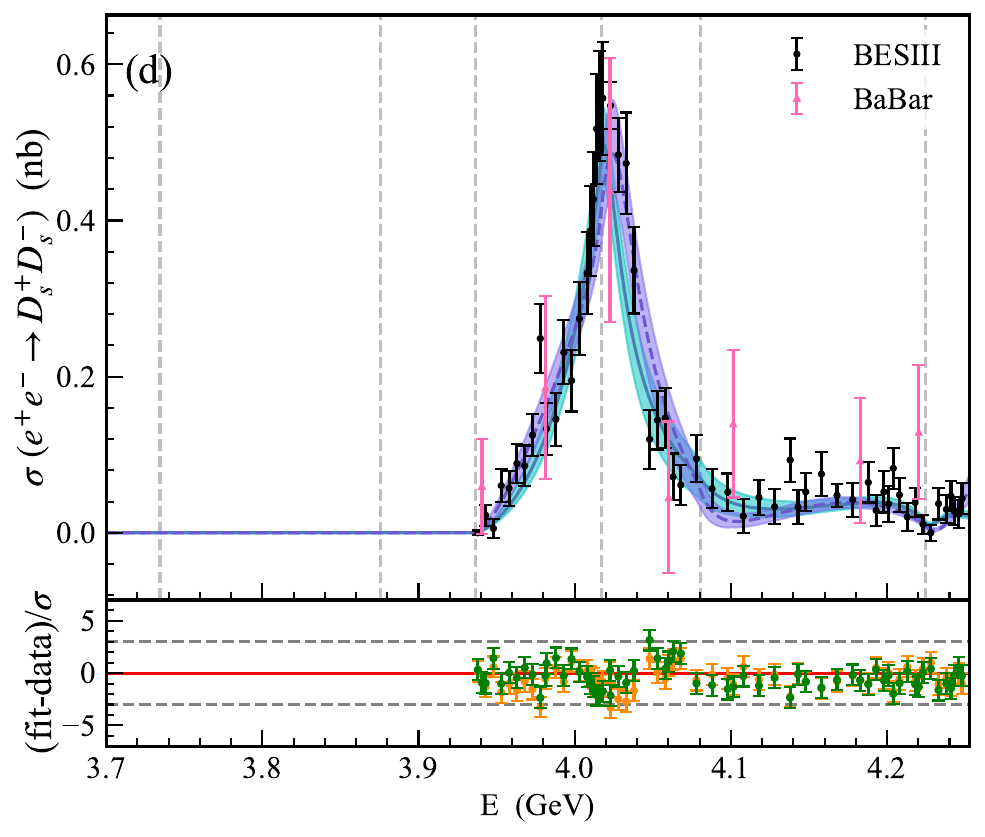}
    \includegraphics[width=0.432\linewidth]{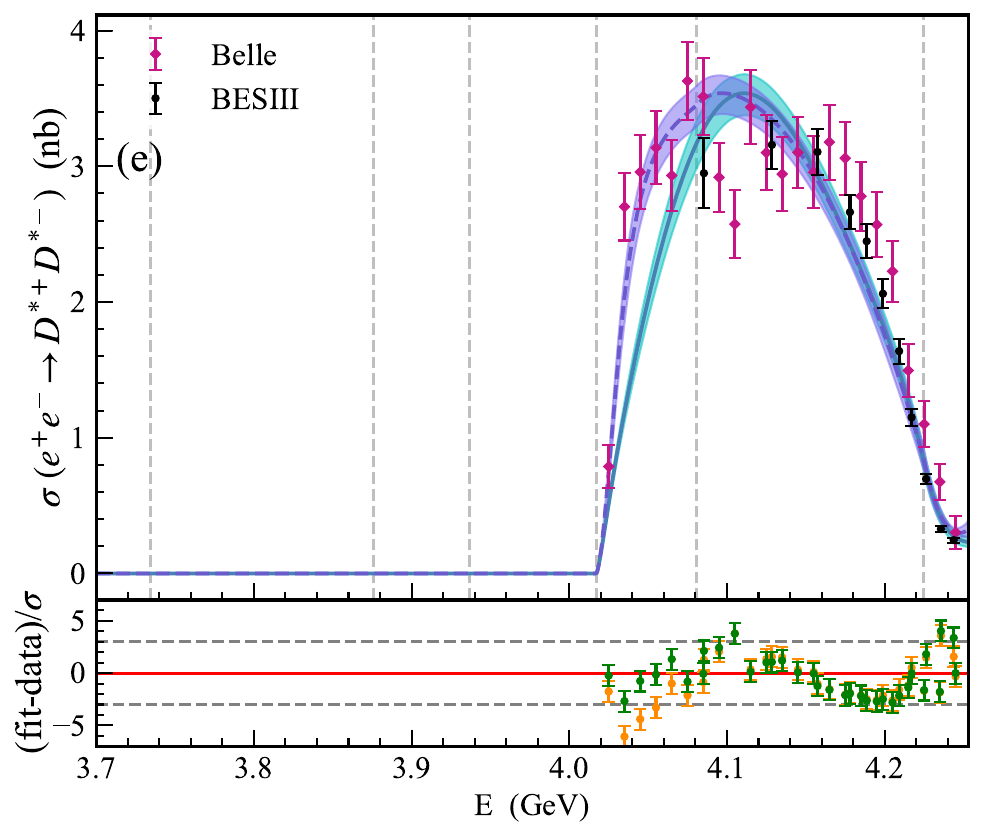}
    \includegraphics[width=0.432\linewidth]{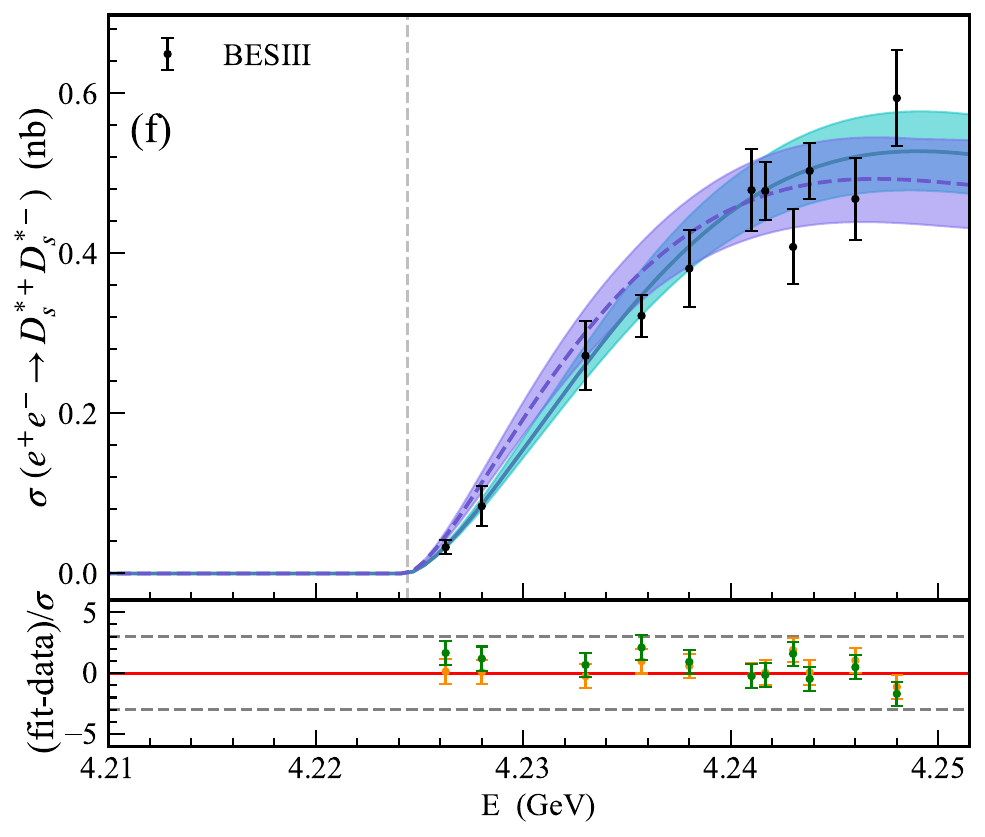}
    \caption{The line shapes of Model I (solid curve) and Model II (dashed curve) in comparison with the experimental data. Panels (a)-(f) show the line shapes of the channels $e^+e^- \to D^+D^-$, $D^0\bar{D}^0$, $D^+D^{*-}$, $D^+_sD^-_s$, $D^{*+}D^{*-}$ and $D^{*+}_sD^{*-}_s$, respectively. The $D\bar{D}$ data is from both BESIII \cite{BESIII:2024ths,julin2017measurement} and Belle \cite{Belle:2007qxm} collaborations. The experimental data in the $D\bar{D}^*$ and $D^*\bar{D}^*$ channels are from BESIII \cite{BESIII:2021yvc} and Belle \cite{Belle:2017grj} collaborations. The $D_s^+D_s^-$ data are from BESIII \cite{BESIII:2024zdh} and BaBar \cite{BaBar:2010plp} collaborations. The data in the $D_s^{*+}D_s^{*-}$ are from BESIII collaboration~\cite{BESIII:2023wsc}. The blue and the purple region denote the $99\%$ confidence levels for Model I and Model II, respectively. The six vertical gray dashed lines represent the $D\bar{D}$, $D\bar{D}^*$, $D_s^+D_s^-$, $D^{*}\bar{D}^*$, $D_s^+D_s^{*-}$ and $D_s^{*+}D_s^{*-}$ thresholds from left to right, respectively. The lower panel of each figure is the standardized residual plot, where the orange and green points represent the standard residuals of Model I and Model II, respectively.}
    \label{line_shape}
\end{figure*}

In this section, we perform a global fit to the latest experimental cross sections of the $e^+e^-\to D\bar{D}$ \cite{Belle:2007qxm,julin2017measurement,BESIII:2024ths}, $e^+e^-\to D^*\bar{D}+c.c.$ \cite{Belle:2017grj,BESIII:2021yvc}, $e^+e^-\to D^*\bar{D}^*$ \cite{Belle:2017grj,BESIII:2021yvc} processes within the energy region $[3.7,4.25]~\mathrm{GeV}$. More specifically, we fit the cross sections of the eight processes $e^+e^-\rightarrow (D\bar{D})^{u,d,s}, (D\bar{D}^*)^{d}, (D^*\bar{D}^*)^{d,s}_{S=0,2}$. The fitting is performed by imimuit \cite{JAMES1975343} with over 1000 starting values to find the global minimum value. The fitted cross sections and the dynamical parameters governing the scattering amplitudes Eq.~(\ref{LSE}) (and thus the pole positions)
of the two models are presented in Fig.~\ref{line_shape} and Tab.~\ref{para}. The other parameters are listed in Table \ref{tab2:parameter} of App.~\ref{FIT}.

\begin{figure}
    \centering
    \includegraphics[width=0.49\linewidth]{./Pole1.pdf}
    \includegraphics[width=0.49\linewidth]{./Pole2.pdf}
    \caption{Pole positions on the complex energy $E$-plane (left) and momentum $k$-plane (right). The red line on the real axis represents the energy region above the threshold. The green solid circle represents a bound state, located below the threshold on the physical RS, the blue hollow triangle denotes a virtual state, located below the threshold on the unphysical RS, and the orange square represents a resonance state, located above the threshold on the unphysical RS, with a non-zero imaginary part.}
    \label{Pole_position}
\end{figure}

\begin{table}
\renewcommand{\arraystretch}{1.5}
    \centering
    \caption{The dynamical parameters governing the scattering amplitudes Eq.~(\ref{LSE}) in the fitting.}
    \begin{tabular}{p{2.6cm}<{\centering}p{2.5cm}<{\centering}p{2.5cm}<{\centering}}
    \hline\hline
   \textbf{Parameters}  &  \textbf{Model I} & \textbf{Model II} \\
    \hline
   $g_{1D}^0~[\mathrm{GeV^{-1}}]$  & $0.66\pm 0.04$    & $-12.93\pm 0.26$  \\
   $g_{3S}^0~[\mathrm{GeV^{-1}}]$  & $-14.66\pm 0.37$  &  $-14.11 \pm 0.96$  \\
   $g_{2D}^0~[\mathrm{GeV^{-1}}]$  & $-17.09\pm 0.23$  & $-$  \\ 
   $m_{1D}^0~[\mathrm{GeV}]$  & $3.807\pm 0.001$ & $3.804\pm 0.001$  \\
   $m_{3S}^0~[\mathrm{GeV}]$  & $4.229 \pm 0.002$ & $4.253\pm 0.005$ \\
   $m_{2D}^0~[\mathrm{GeV}]$  & $3.692\pm 0.003$  & $-$ \\
   $\Lambda~[\mathrm{GeV}]$ & $0.50 \pm 0.00$ & $0.50\pm 0.00$\\
   $\chi^2/\mathrm{d.o.f.}$ & 2.17  &  2.66\\
   \hline
   \hline
   \end{tabular} 
   \label{para}
\end{table}

\begin{figure}[hbt!]
    \centering
    \includegraphics[width=1\linewidth]{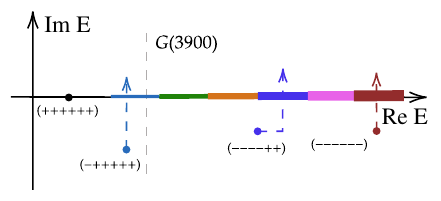}
    \caption{Paths from poles in Model I on unphysical RSs to the physical RS. Poles are represented by solid circles. The blue, green, orange, dark blue, purple and red solid line denote the cuts form the $D\bar{D}$, $D\bar{D}^*$, $D_s^+D_s^-$, $D^*\bar{D}^*$, $D_s^+D_s^{*-}$ and $D_s^{*+}D_s^{*-}$ thresholds, respectively.}   
    \label{path}
\end{figure}

From Fig.~\ref{line_shape}, one can see significant contributions of the $\psi(3770)$ in the $D^+D^-$ and $D^0\bar{D}^0$ channels. 
The signals of the $\psi(4040)$ in the $D^+D^{*-}$ and $D_s^+D_s^-$ channels are more pronounced than that in the $D\bar{D}$ channel. 
Whether the vector chamronium-like state $G(3900)$ exists or not needs further pole analysis. As shown in the standardized residual plot in Fig.~\ref{line_shape}, a large proportion of the standardized residuals are distributed within the interval [-3, 3]. Figures (a), (b), (e), and (f) show that the standardized residuals are approximately randomly distributed around zero, suggesting that the fit is satisfactory. However, there are more standardized residuals below zero than above in Figures (c) and (d), a similar results also observed in Ref.~\cite{Husken:2024hmi}. This result can be attributed to two main reasons. First, there exist significant discrepancies between the data points of BESIII and Belle collaborations in certain regions for the $e^+e^-\to D^+D^{*-}$ and $e^+e^-\to D^{*+}D^{*-}$ processes, which reduces the quality of the fit and prevents the standardized residuals from being randomly distributed around zero. Secondly, under the consideration of coupled-channel effects, there is still a deviation from the experimental data in the $e^+e^-\to D^+D^{*-}$ and $e^+e^-\to D^{*+}D^{*-}$ channels for both  the LSE and the $K$-matrix approaches. The projection of all standardized residuals onto the vertical axis yields a distribution that closely resembles a Gaussian distribution which are presented in Figs.~\ref{SR15} and \ref{SR20} in App.~\ref{residual}. The standardized residuals of Model I show a slight deviation from the Gaussian distribution, which may indicate mild overfitting. In comparison, the residuals of Model II are more consistent with a Gaussian distribution, suggesting a higher quality of fit. Overall, the fitting performance of both models remains within an acceptable range. 

In Tab.~\ref{para}, the parameters $g_{1D}^0$, $g_{3S}^0$ and $g_{2D}^0$ denote the bare couplings between the charmonia $\psi(1D)$, $\psi(3S)$ and $\psi(2D)$ and an open charmed meson pairs. $m_{1D}^0$, $m_{3S}^0$ and $m_{2D}^0$ represent the bare mass of $\psi(1D)$, $\psi(3S)$ and $\psi(2D)$, respectively. $\Lambda$ is the cutoff parameter in the two-point loop function. The reduced chi-squares are $\chi^2/\mathrm{d.o.f}=2.17$ and $\chi^2/\mathrm{d.o.f}=2.66$
for Model I and Model II, respectively, which indicates that an additional bare vector charmonium state greatly optimizes the fit result. With the fitted parameters, we can extract the  physical quantities of interest, such as pole positions in the complex $E$-plane, the effective couplings, and so on. With the fitted parameters, we can extract the  physical quantities of interest, such as pole positions in the complex $E$-plane, the effective couplings, and so on.

\subsection{Pole analysis}

A state is identified as a pole of the $T$-matrix in the complex energy plane, either bound state, virtual state, or resonance (Fig.~\ref{Pole_position}). The pole positions can be obtained by solving the equation 
\begin{equation}
\textbf{det}\left[1-\hat{V}_{\text{oo}}^{\text{eff}}G_{CT}(E_r)\right]=0~.
\end{equation}
Through analytic continuation, the complex $E$-plane can be extended to $2^n$ (with $n$ the number of involved channels) Riemann sheets (RSs) which are labeled by $(\pm,\dots,\pm)$ according to the signs of the imaginary parts of the c.m. three-momenta in the two-body channels. The physical RS is denoted by $(+,\dots,+)$, whereas the others represent the unphysical RSs. The physical and unphysical RSs are connected by branch cuts where the two-point function satisfies
\begin{align}
    G^{\uppercase\expandafter{\romannumeral2}}_{ii}(E-i\varepsilon)=G^{\uppercase\expandafter{\romannumeral1}}_{ii}(E+i\varepsilon).
\end{align}
Here, the indices \uppercase\expandafter{\romannumeral1} and \uppercase\expandafter{\romannumeral2} represent the two-point functions on the first (physical) and second (unphysical) RSs, and $i=1,\dots,n$ denote the $i$th channel.

We assume that the $D^{(*)+}$ meson mass is equal to the $D^{(*)0}$ meson mass,  because its mass difference is very small. As the result,  it is unnecessary to search for poles in the region $[D^{0}\bar{D}^{0}]_{\text{Thr}}<E_r<[D^{+}D^{-}]_{\text{Thr}}$, $[D^{0}\bar{D}^{*0}]_{\text{Thr}}<E_r<[D^{+}D^{*-}]_{\text{Thr}}$ and $[D^{*0}\bar{D}^{*0}]_{\text{Thr}}<E_r<[D^{*+}D^{*-}]_{\text{Thr}}$, where $[(D^{(*)}\bar{D}^{(*)})_{u,d}]_{\text{Thr}}$ represent the threshold of the $(D\bar{D})_{u,d}$ meson pair. Therefore, there are six thresholds in our coupled-channel system, i.e. $[D\bar{D}]_{\text{Thr}}$, $[D\bar{D^*}]_{\text{Thr}}$, $[D_s^+\bar{D}_s^-]_{\text{Thr}}$, $[D^*\bar{D}^*]_{\text{Thr}}$, $[D_s^+\bar{D}_s^{*-}]_{\text{Thr}}$ and $[D_s^{*+}\bar{D}_s^{*-}]_{\text{Thr}}$ in order, where we use $D^{(*)}$ to denote $D^{(*)+}$ and $D^{(*)0}$ mesons. In the following, we use the abbreviations $\text{Thr}_1$, $\text{Thr}_2$,$\dots$, $\text{Thr}_6$ to denote the six thresholds. 
In general, we need to find the poles on $2^6$ RSs. However, in practice, we are only concerned with the poles on the physical RS $(+,+,+,+,+,+)$ and those on the unphysical RSs close to the physical region. These unphysical RSs are labeled by sequentially replacing the plus signs with minus signs, i.e. $(-,+,\dots,+)$, $(-,-,\dots,+)$,$\dots$, $(-,-,\dots,-)$.
Therefore, we find poles for the following cases:
\begin{enumerate}
    \item Poles below all the thresholds on the physical RS $(+,+,+,+,+,+)$.
    \item Poles above the $i$-th threshold and below the $(i+1)$-th threshold on the unphysical RS $(-,,\dots, -_i,+_{(i+1)},\dots,+)$.
    \item Poles below but close to the $i$-th threshold on the unphysical RS $(-,,\dots, -_i,+_{(i+1)},\dots,+)$, which also have impact on the physical observables.  
\end{enumerate}
Whether the above poles have a significant physical impact or not depends on their distance $E_d$ to the physical RS. 
 In Fig.~\ref{path}, we present a schematic diagram of the distance $E_d$ of the poles on different unphysical RSs. Although the poles of the resonance states are symmetric about the real axis, only the poles on the lower half-plane are close to the physical RS as shown in Fig.~\ref{path}.
 For Case 1, since the pole is already on the physical RS, the distance to the physical sheet is zero. For Case 2, since the unphysical RS $(-,\dots, -_i,+_{(i+1)},\dots,+)$ is connected to the physical RS $(+,+,+,+,+,+)$ along the region $[\text{Thr}_i,\text{Thr}_{i+1}]$ on the real axis, where $\text{Thr}_i$ denotes the $i$-th threshold. Those poles can reach to the physical RS by moving the energy distance along the direction of positive imaginary axis, where $E_d$ is equal to the modulus of the imaginary part of the pole. Finally, for Case 3, the pole can firstly move $E_r$ along the direction of positive real axis, with $E_r$ the difference between the threshold and the real part of the pole. Furthermore, they move along the direction of positive imaginary axis as that for Case 2. Thus, $E_d$ is $E_r$ plus the modulus of the imaginary part of the pole. The poles in Model~I and Model~II are presented in Tab.~\ref{pole_I}. The effective couplings of the poles to the $i$th-channel in model I and model II are listed in Tables \ref{tab:1nnModelI} and \ref{tab:1nnModelII} of App.~\ref{FIT}, respectively. In Tab.~\ref{pole_I}, one can see the two poles corresponding to the $\psi(3770)$ and $\psi(4040)$ in both Model I and Model II. 
\begin{table}
\renewcommand{\arraystretch}{1.5}
    \centering
    \caption{Pole positions on the various RSs. The numbers in square brackets represent energy distances of the poles to the physical RS, in units of $\mathrm{MeV}$. }
    \begin{tabular}{ccc}
        \hline
        \hline
        \textbf{Riemann sheets} & \textbf{Model I} & \textbf{Model II}  \\
        \hline
        $(+,+,+,+,+,+)$ & 3.691.60 & $-$ \\
        \hline
        \multirow{2}{*}{$(-,+,+,+,+,+)$} & $-$ & $3743.07 \pm 7.36i$ [7] \\
        & $3778.42\pm 11.81i$ [12] & $3775.29 \pm 14.31i$ [14] \\
        \hline
        $(-,+,-,+,+,+)$ & $3832.52\pm 74.53i$ & $-$ \\
        $(-,-,+,+,+,+)$ & $-$ & $3883.91\pm46.53i$ [47] \\
        $(-,-,-,-,+,+)$ & $4011.05\pm 10.13i$ [16] & $4019.42 \pm 17.40i$~[17] \\
        $(-,-,-,-,-,-)$ & $4232.78\pm 23.96i$ [24] & $4278.21\pm 21.59i$ [22] \\
        \hline
        \hline
        \end{tabular}
   \label{pole_I}
\end{table}
 In Model I, a pole $3832.52\pm 74.53i$ around $3.9~\mathrm{GeV}$ on the $(-,+,-,+,+,+)$ contributes to the broader peak structure in experiment. In Model II, a pole $3883.91\pm 46.53 i$ on the $(-,-,+,+,+,+)$ sheet corresponds to the peak structure around $3.9~\mathrm{GeV}$.
\begin{figure*}
    \centering
    \includegraphics[width=0.45\linewidth]{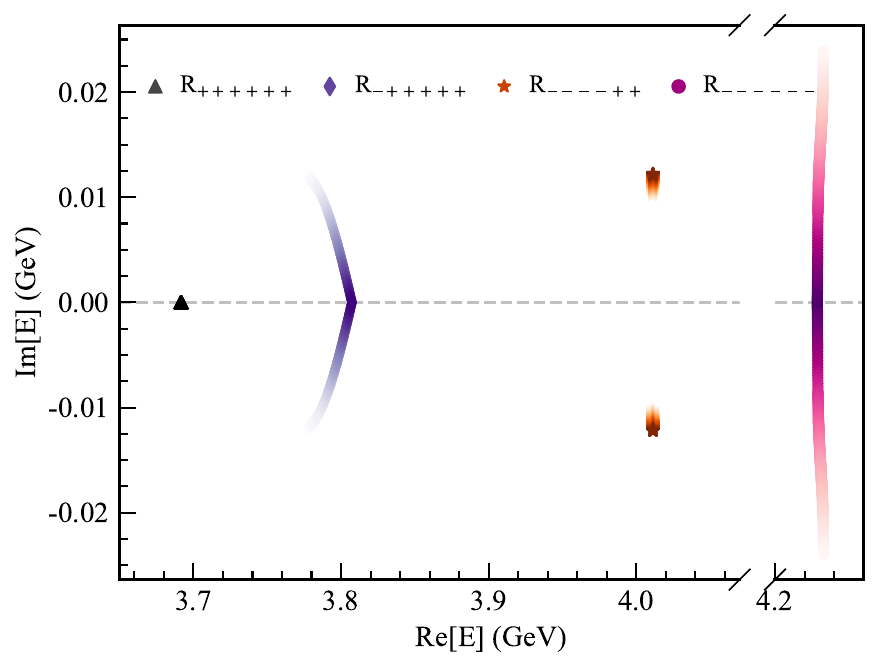}
    \includegraphics[width=0.45\linewidth]{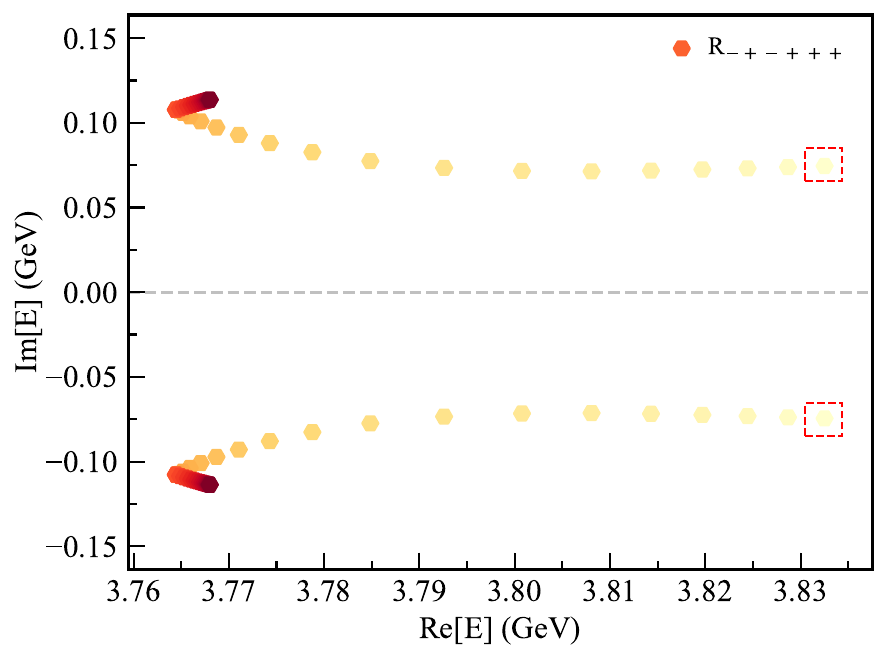}
    \caption{Left: The trajectories of the poles in Model~I on various RSs with the coupling constants $g_{2D}^0$, $g_{1D}^0$ and  $g_{3S}^0$ varying sequentially from the fitted values to zero. Right: The trajectory of pole $3832.57^{+0.91}_{-0.79}\pm 74.53^{+0.68}_{-2.15}i$ MeV on the $(-,+,-,+,+,+)$ RS. Different shapes represent the trajectories of distinct poles. The red dashed rectangular boxes represent the initial positions of the poles and the colors of the poles gradually become darker as the parameters $g_{2D}^0$, $g_{1D}^0$ and $g_{3S}^0$ vary sequentially from the fitted values to zero.}   
    \label{trajectory}
\end{figure*}

However, so far, we are unable to directly determine which poles are dynamically generated states and which are from a strong renormalization of bare vector charmonia. 
One can distinguish them by plotting the trajectory of the poles when the coupling constants vary. 
 Since the procedures for calculating the pole trajectory are similar between the two models, only the pole trajectories for Model~I are presented. The specific procedure is as follows:
\begin{enumerate}
    \item  Vary the coupling constant $g_{2D}^0$ from the fitted value to $0$  with $g_{1D}^0$ and $g_{3S}^0$ fixed to their fitted values and plot the trajectory of all the poles.
    \item Vary the coupling constant $g_{1D}^0$ from the fitted value to $0$ with $g_{3S}^0$ and $g_{2D}^0$ fixed to the fitted value and $0$, respectively. Plot the trajectory of all the poles.
    \item Perform the same procedure for the coupling constant $g_{3S}^0$ with the couplings $g_{2D}^0$ and $g_{1D}^0$ fixed to $0$. Plot the trajectory of all the poles as $g_{3S}^0$ varies.
\end{enumerate}
When the coupling constants gradually change to zero, if the pole moves towards the bare mass obtained from the fitting on the real axis, it indicates that the pole is a state renormalized from a bare state. For the sake of convenience, we illustrate the reason for the case of one open charmed channel and one bare charmonium state. In this case, the $T$-matrix reads as
\begin{align}
    T=\frac{v_{CT}+\frac{g^2}{s-m^2}}{1-(v_{CT}+\frac{g^2}{s-m^2})G(s)},
\end{align}
where $v_{CT}$ and $g$, which are both real numbers, are the contact potential and the coupling between the bare charmonium and open charmed meson pair. The pole $s_0$ is the solution of $1-(v_{CT}+\frac{g^2}{s-m^2})G(s)=0$. When $g=0$, one obtains
\begin{align}
    1-v_{CT}G(s_0)=0.
\end{align}
In this case, $\sqrt{s_0}$ corresponds to the dynamically generated state, and the contribution of any charmonium vanishes. When $g\neq 0$, we can extract the pole from
\begin{align}
    s_0= \frac{g^2G(s_0)}{1-v_{CT}G(s_0)}+m^2.
\end{align}
When $g$ gradually decreases to zero, the term $\frac{g^2G(s_0)}{1-v_{CT}G(s_0)}\rightarrow 0$. Therefore, the pole $\sqrt{s_0}$ will move towards the bare mass of the charmonium. Similarly, for the system with many open charmed channels and many bare charmonia, this property still works. As long as we gradually adjust all the couplings to zero, the renormalized charmonium state will move to its bare mass. 
The pole trajectories in Model~I are presented in Fig.~\ref{trajectory} and the zoomed-in diagram of all poles in the left diagram can be found in App.~\ref{all_tra}. 

\begin{figure}
    \centering
    \includegraphics[width=0.9\linewidth]{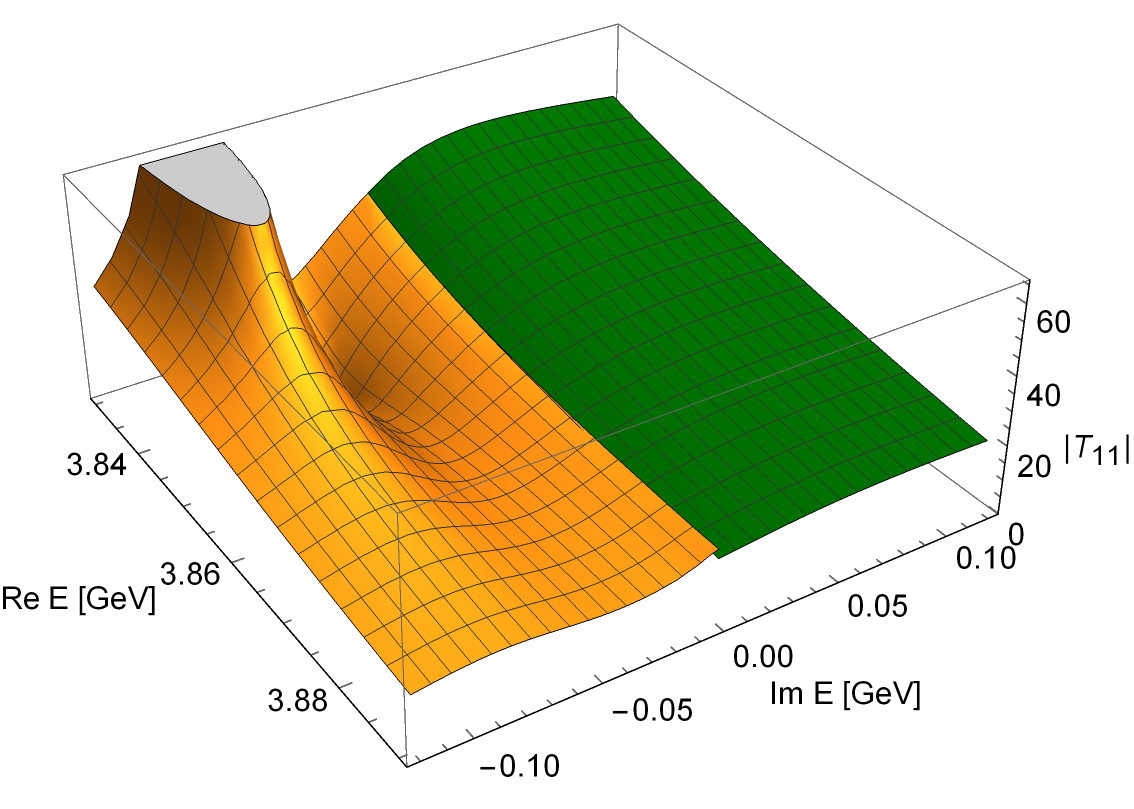}
    \caption{The modulus of the scattering amplitude $|T_{11}|$ on the complex $E$-plane. The orange and green surfaces are the lower half plane of the $(-,+,-,+,+,+)$ RS and the upper half plane of physical RS, respectively.}   
    \label{T_11}
\end{figure}

The pole trajectories of Model~I are illustrated in Fig.~\ref{trajectory}. When the couplings $g_{2D}^0$, $g_{1D}^0$ and $g_{3S}^0$ vary from their fitted values to zero, the $3691.60~\mathrm{MeV}$, $3778.42\pm 11.81i~\mathrm{MeV}$ and $4232.78\pm 23.96i~\mathrm{MeV}$ poles approach to the bare masses $m_{2D}^0$, $m_{1D}^0$ and $m_{3S}^0$, respectively. These three poles are considered as the $\psi(2D)$, $\psi(1D)$ and $\psi(3S)$ vector charmonia. The significant discrepancy between the $3691.60$~MeV pole and the experimental result indicates that the  $\psi(2D)$ charmonium acts as a redundant free parameter in the fit without any physical significance. This suggests that an additional charmonium $\psi(2D)$ is unnecessary, even though it has improved the fitting results a little bit. On the contrary, the pole $4011.05 \pm 10.13i$ MeV undergoes only slight movement on the RSs which is considered as a dynamically generated state. 

\begin{figure}
    \centering
\includegraphics[width=0.95\linewidth]{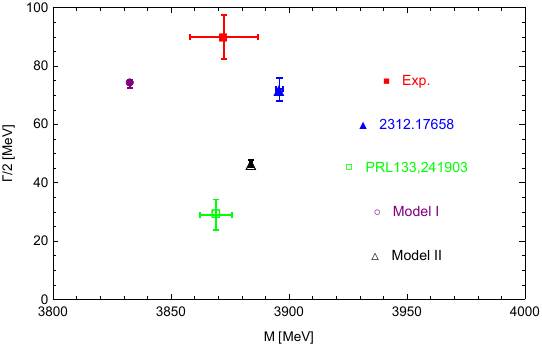}
    \caption{The pole positions of Model I (purple hollow circle) and Model II (black hollow triangle) in comparison with other works. Blue triangle and green hollow box are the results from Ref.~\cite{Nakamura:2023obk} and Ref.~\cite{Lin:2024qcq}, respectively. Red box is the experimental result~\cite{BESIII:2024ths}.}   
    \label{fig:pole}
\end{figure}

After carefully searching for poles on other RSs, we find another dynamically generated state at $3832.57^{+0.91}_{-0.79}\pm 74.53^{+0.68}_{-2.15}i$~MeV, about 40~MeV below the $[D\bar{D}^*]_{\text{Thr}}$, on $(-,+,-,+,+,+)$ sheet. To check its impact on the physical observables, we plot the three-dimensional representation of $|T_{11}|$ which is the modulus of the first row and first column element of $T$-matrix on both $(-,+,-,+,+,+)$ and the physical RS. From Fig.~\ref{T_11}, the curvature of $|T_{11}|$ on the physical plane is nearly the same as that on the $(-,+,-,+,+,+)$ sheet, which indicates that despite the $(-,+,-,+,+,+)$ sheet being relatively far from the physical RS, it still exerts a significant influence on the physical region. 
\begin{table}
\renewcommand{\arraystretch}{1.5}
    \centering
    \caption{The pole position of the $G(3900)$, in comparison with other works. Exp. represents the experimental data from Ref.~\cite{BESIII:2024ths} (Supplemental materials). I and II stand for our Model I and Model II results, respectively.}
    \begin{tabular}{p{2.5cm}<{\centering}p{5cm}<{\centering}}
    \hline\hline
     \multirow{2}{*}{This work} & I~$3832.6_{-0.8}^{+0.9}-74.5_{-2.2}^{+0.7}i$ \\ & II~$3883.9_{-0.5}^{+0.4}-46.5_{-1.2}^{+1.2}i$  \\
     \hline
    Ref.~\cite{Husken:2024hmi} &$-$\\
    Ref.~\cite{Lin:2024qcq} & $3869.2_{-6.7}^{+6.7}-29.0^{+5.2}_{-5.2}i$\\
    Ref.~\cite{Nakamura:2023obk} &$3896.0_{-1.4}^{+1.4}-72.0_{-3.9}^{+3.9}i$\\
    Exp. &$3872.5^{+14.2+3.0}_{-14.2-3.0}-89.9^{ +7.0 +2.5}_{-7.0-2.5}i$\\
   \hline
   \hline
   \end{tabular}  
   \label{dif_results}
\end{table}
\begin{table*}[hbt!]
\renewcommand{\arraystretch}{1.5}
    \centering
    \caption{The pole positions of $\psi(3770)$ and $\psi(4040)$, in comparison with the results of other works. I and II stand for Model I and Model II.}
    \begin{tabular}{p{2cm}<{\centering}p{3cm}<{\centering}p{3cm}<{\centering}p{3cm}<{\centering}p{3cm}<{\centering}}
    \hline\hline
     &\multicolumn{2}{c}{$\psi(3770)$}  &\multicolumn{2}{c}{$\psi(4040)$}  \\
     \hline
    This work & I~$3778.42-11.81i$ & II~ $3775.29-14.31i$ & I~$4011.05-10.13i$ & II~ $4019.42-17.40i$ \\
    Ref.~\cite{Husken:2024hmi} & \multicolumn{2}{c}{$3778.7_{-0.7}^{+0.7}-17.0_{-0.2}^{+0.2}i$} & \multicolumn{2}{c}{$4044.0_{-1.5}^{+1.5}-65.0_{-1.5}^{+1.5}i$}\\
    Ref.~\cite{Lin:2024qcq} & \multicolumn{2}{c}{$3778.0^{+0.3}_{-0.3}-12.3^{+0.3}_{-0.3}i$} & \multicolumn{2}{c}{$4019.5_{-0.5}^{+0.5}-22.9^{+1.1}_{-1.1}i$}\\
    Ref.~\cite{Nakamura:2023obk} & \multicolumn{2}{c}{$3780.0_{-1.3}^{+1.3}-15.2_{-1.1}^{+1.1}i$} & \multicolumn{2}{c}{$4029.2_{-0.4}^{+0.4}-14.0_{-0.5}^{+0.5}i$}\\
    PDG & \multicolumn{2}{c}{$3773.7_{-0.7}^{+0.7}-13.6_{-0.5}^{+0.5}i$} & \multicolumn{2}{c}{$4039.6_{-4.3}^{+4.3}-42.3_{-6.2}^{+6.2}i$}\\
   \hline
   \hline
   \end{tabular}  
   \label{Other_pole}
\end{table*}
\begin{table*}
\renewcommand{\arraystretch}{1.5}
    \centering
    \caption{Poles positions and effective couplings of the $1^{-+}$ system in Model I on different RSs. The dimension of coupling is $\mathrm{GeV}^{-3/2}$. The square brackets represent energy distance the poles move to the physical RS. The unit is $\mathrm{MeV}$. The effective couplings with underline are the largest couplings for a given pole, which indicate the dominant channel.} 
    \begin{tabular}{p{3cm}p{3cm}<{\centering}p{2cm}<{\centering}p{2cm}<{\centering}p{2cm}<{\centering}p{2cm}<{\centering}}
    \hline\hline 
     \textbf{Riemann Sheets} &\textbf{Poles} $[\mathrm{MeV}]$  &$g_{D\bar{D}^*}$  &$g_{D^*\bar{D}^*}$ & $g_{D_s^+D_s^{*-}}$ & $g_{D_s^{*+}D_s^{*-}}$ \\
    \hline
    $(+,+,+,+)$ & 3836.57 & 8.59 & \underline{32.04} & 0.04 & 0.14  \\
    $(-,+,+,+)$ & $3885.42\pm 9.48i$ [10] & 2.21 & 6.70 & 7.66 & \underline{29.46}  \\
    $(-,-,+,+)$ & $4001.56\pm 3.94i$ [19]& 0.31 & \underline{1.50} & 0.01 & 0.03  \\
    $(-,-,-,+)$ & $4085.70\pm 27.08i$ [27]& 0.13 & 0.42 & 2.25 & \underline{6.75}  \\
    $(-,-,-,-)$ & $4224.18\pm31.26i$ [31]& 0.04 & 0.08 & 0.50 & \underline{1.99}  \\
   \hline
   \hline
   \end{tabular}  
   \label{1npI}
\end{table*}
\begin{table*}
\renewcommand{\arraystretch}{1.5}
    \centering
    \caption{Poles positions and effective couplings of the $1^{-+}$ system in Model II on different RSs. Other details are similar to Tab.~\ref{1npI}.}
    \begin{tabular}{p{3cm}p{3cm}<{\centering}p{2cm}<{\centering}p{2cm}<{\centering}p{2cm}<{\centering}p{2cm}<{\centering}}
    \hline\hline
     \textbf{Riemann Sheets} &\textbf{Poles} $[\mathrm{MeV}]$  &$g_{D\bar{D}^*}$  &$g_{D^*\bar{D}^*}$ & $g_{D_s^+D_s^{*-}}$ & $g_{D_s^{*+}D_s^{*-}}$ \\
    \hline
    $(+,+,+,+)$ & 3869.57 & 4.38 & \underline{8.19} & 0.02 & 0.09  \\
    $(-,+,+,+)$ & $3891.73\pm 26.19i$ [26] & 1.77 & 13.68 & 0.92 & \underline{39.25}  \\
    $(-,-,+,+)$ & $4017.93\pm 2.71i$ [3]& 0.21 & \underline{2.34} & 0.01 & 0.04  \\
    $(-,-,-,+)$ & $4087.76\pm 21.92i$ [22]& 0.18 & 0.30 & 2.35 & \underline{12.02}  \\
    $(-,-,-,-)$ & $4213.85\pm9.63i$ [20]& 0.07 & 0.21 & 0.40 & \underline{2.11}  \\
   \hline
   \hline
   \end{tabular}  
   \label{1npII}
\end{table*}

For Model~II, the poles and the corresponding effective couplings are presented in App.~\ref{FIT}. Through a pole trajectory analysis similar to that of Model~I, the poles $3775.29\pm 14.31i$ MeV and $4278.21\pm 21.59i$ MeV approach to the bare masses $m_{1D}^0$ and $m_{3S}^0$, corresponding to the $\psi(1D)$ and $\psi(3S)$ charmonia, respectively. The other poles, interpreted as dynamically generated states, remain largely stationary as the couplings $g^0_{1D}$ and $g^0_{3S}$ change from their fitted values to zero. In contrast to Model I, a new pole $3743.07\pm 7.36i$ MeV emerges on $(-,+,+,+,+,+)$ sheet in Model II. This state does not manifest itself as a visible peak in the $e^+e^-\to D\bar{D}$ cross section, possibly because it couples weakly to the $D\bar{D}$ channel and lies close to $\psi(3770)$. Alternatively, shown, the pole couples predominantly to the $D^*\bar{D}^*$ channel instead of the $D\bar{D}$ channel (App.~\ref{FIT}). Consequently, its contribution to the $e^+e^-\to D\bar{D}$ cross section is suppressed, and further obscured by the overlapping of the nearby $\psi(3770)$, making it difficult to be observed in experiment. The pole $3883.91^{+0.38}_{-0.46}\pm 46.53^{+1.22}_{-1.22}i$~MeV, 9~MeV above the $[D\bar{D}^*]_{\text{Thr}}$ threshold,  locates on the $(-,-,+,+,+,+)$ sheet and can be considered as a candidate of the $G(3900)$.  The central value of the real part is consistent with the experimental mass of the $G(3900)$ within the uncertainty. The width of this pole is considerably narrower than that obtained by BESIII, suggesting that the Breit-Wigner fit used by BESIII may not be suitable for describing a near-threshold state. 

In Model I, we identify the pole located at $3832.6^{+0.9}_{-0.8}-74.5_{-2.2}^{+0.7}i$ MeV as the candidate of the $G(3900)$. 
The central value of the real part is approximately 40~MeV lower than the mass of the $G(3900)$. While its width is consistent with the experimental large value, making it still a visual broad structure in experiment. Although the real part of the pole lies below $[D\bar{D}^*]_{\text{Thr}}$, the state can still decay into the $D\bar{D}^*$ final state, which makes it a plausible candidate for the $G(3900)$.

The comparison between our results and  other works~\cite{Husken:2024hmi,Lin:2024qcq,Nakamura:2023obk} is presented in Tab.~\ref{dif_results} and Fig.~\ref{fig:pole}. 
Ref.~\cite{Husken:2024hmi} uses the K-matrix parametrization to describe the cross sections of the $e^+e^-\to D^{(*)}\bar{D}^{(*)}$ processes, but without hidden strange channels. 
Their overall fitting indicates that the $G(3900)$ is only a threshold enhancement of the $D^*\bar{D}$ channel. Ref.~\cite{Lin:2024qcq} uses the One-Boson-Exchanged (OBE) model to connect the dynamics of the $S$-wave hadronic molecule $\chi_{c1}(3872)$, 
$Z_c(3900)$, $T_{cc}(3875)$ 
to the $P$-wave $G(3900)$. 
The proper $S$-wave pole positions indicate the existence of the $P$-wave hadronic molecule $G(3900)$. Ref.~\cite{Nakamura:2023obk} obtains  the same conclusion by fitting to 18 two-body and three-body hadronic channels, which also needs to deal with the three-body dynamics properly.    
 
 The results obtained from both Model I and Model II are in agreement with those reported in Ref.~\cite{Lin:2024qcq,Nakamura:2023obk}, supporting the interpretation of $G(3900)$ as a $P$-wave $D\bar{D}^*/\bar{D}D^*$ molecule state.
While the inclusion of the bare $\psi(2D)$ state in Model I introduces three additional parameters, potentially causing overfitting in Model I. Given that $\chi^2/\mathrm{d.o.f}$ of Model II is already satisfactory and that its residue analysis shows better performance, we place greater emphasis on the poles found in Model II. Although both models yield consistent conclusions regarding the dynamical origin of the pole, we adopt the pole positions from Model II due to its good residue performance (the lower panel of Fig.~\ref{line_shape}). 

Similarly, we also compare the results for $\psi(3770)$ and $\psi(4040)$ with those reported in Refs.~\cite{Husken:2024hmi,Lin:2024qcq,Nakamura:2023obk}, presented in Tab.~\ref{Other_pole}. We note that the pole positions of the $\psi(3770)$ extracted from Models I and II are in good agreement with the results reported in Refs.~\cite{Husken:2024hmi,Lin:2024qcq,Nakamura:2023obk} and the value listed by the PDG. In this work, the pole position of the $\psi(4040)$ is essentially consistent with that in Ref.~\cite{Lin:2024qcq}, and its width is smaller than the value reported by the PDG, which suggests the Breit-Winger fit may not be suitable for describing a near-threshold state. Compared with Refs.~\cite{Husken:2024hmi,Lin:2024qcq,Nakamura:2023obk}, we note that although all these studies incorporate a bare state $\psi(4040)$, Refs.~\cite{Husken:2024hmi,Lin:2024qcq,Nakamura:2023obk} do not investigate whether the pole originates from the bare state or a dynamic state. Based on our tests, we find that within our framework, $\psi(4040)$ is a dynamically generated state, and the bare state we introduce is shifted to $4232.78-23.96i$~MeV for Model I and $4278.21-21.59i$ MeV for Model II. One possible explanation for this shift is the presence of high thresholds in the coupled-channel dynamics, which may have significantly effect on the bare state positions. 

In a short summary, our analysis indicates that the $G(3900)$ originates from a dynamically generated pole, consistent with the conclusions of  Ref.~\cite{Lin:2024qcq,Nakamura:2023obk}. Although Model II yields a slightly larger $\chi^2/\text{d.o.f}$ compared to Model I, the potential overfitting in Model I caused by additional parameters leads us to place greater emphasis on the results of Model II. This suggests that two bare states are sufficient to describe the experimental data in the energy region $[3.7,4.25]~\mathrm{GeV}$. Additionally, we extract the differential cross section (presented in Fig.~\ref{Dis} of App.~\ref{all_tra}) of the $D\bar{D}$ channel at $\sqrt{s}=3.873~\mathrm{GeV}$, which is the experimental mass position of the $G(3900)$.

 At last, we stress that the $D^{(*)}\bar{D}^{(*)}$ system with quantum number $J^{PC}=1^{--}$ in a $P$-wave configuration presents challenges within the framework of effective field theories (EFTs). Such $P$-wave interactions in hadronic systems are claimed to induce non-trivial renormalization behavior by some authors~\cite{Nogga:2005hy,Rotureau:2010uz}. Specifically, it is argued that the implementation of consistent power counting requires the introduction of higher-order counter terms to preserve renormalization group invariance, particularly when dealing with singular potentials characteristic of $P$-wave interactions. However, this point of view is challenged in Refs.~\cite{Epelbaum:2006pt,Epelbaum:2018zli,Epelbaum:2020maf,Epelbaum:2021sns}, where it is shown that a self-consistent and practically applicable solution to the problem of non-perturbative renormalization is provided by the cutoff EFT. 
In the present work, we intentionally circumvent this debate by performing a phenomenological study as we focus on pole extraction rather than pursuing a complete renormalization procedure. The description of the data thus relies on the choice of the cutoff as a consequence of omitting these necessary counter terms. The phenomenological approach used in this work remains justified for our primary objective of identifying and characterizing possible pole structures in the complex energy plane, while acknowledging that a more fundamental EFT treatment would require systematic inclusion of higher-order counter terms to achieve proper renormalization. The extracted pole positions, while regulator-dependent in technical terms, nevertheless still provide crucial physical insights into the possible existence and qualitative features of exotic hadronic states in this channel.

\subsection{Searching the $1^{-+}$ exotic candidate}

As discussed in the previous section, the dynamics of the $J^{PC}=1^{-+}$ channel is described by the same parameter set as that of the $J^{PC}=1^{--}$ channel. We can also extract the pole positions 
 by plugging the fit parameters into Eq.\eqref{T_1np} and solving \textbf{det} $[\bm{1}_{6\times 6}-V_{CT}^{1-+}G_{CT}(E)]=0$. 
The only relevant parameters are $C_1^i$ and $C_3^i$ due to the appearance of the $|0\times 1\rangle^i$ and $|1\times 1\rangle ^i$ components and the bare charmonium state is absent in this channel. There are four thresholds in the $1^{-+}$ system, i.e. $[D\bar{D}^*]_{\text{Thr}}$, $[D^*\bar{D}^*]_{\text{Thr}}$, $[D_s^+D_s^-]_{\text{Thr}}$ and $[D_s^{*+}D_s^{*-}]_{\text{Thr}}$. The pole positions on the $E$-plane of the $1^{-+}$ channel and their effective couplings to all the channels are presented in Tab.~\ref{1npI} and Tab.~\ref{1npII}, respectively.
For the effective couplings, the $T$-matrix exhibits the following behavior
\begin{align}
    T_{ij} \sim \frac{g_ig_j}{E-E_r}
\end{align}
around the pole $E_r=M_r-i\Gamma_r/2$, with $M_r$ and $\Gamma_r$ the mass and width of a given state. Here, $g_i$ is the coupling of the state to the $i^{\rm th}$-channel. $g_i$ is generally a complex number, and its modulus is conventionally used to represent its magnitude. The coupling constant $g_i$ is obtained from the residues of the $T$-matrix via
\begin{align}
    g_ig_i=\lim_{E\to E_r}(E-E_r)T_{ii}(E),
\end{align}
where $T_{ii}$ can be obtained by Eq.~\eqref{T_matrix}. 

Since the $1^{-+}$ system does not contain bare states and only has contact interactions, all the poles are dynamically generated states. In Model~I, we find a bound state at $3836.57$~MeV on the physical RS, which is around $38$~MeV below the $[D\bar{D}^*]_{\text{Thr}}$. There are also other four resonances at $3885.42\pm 9.48i$~MeV, $4001.56\pm 3.94i$~MeV, $4085.70\pm 27.08i$~MeV, and $4224.18\pm 31.26i$~MeV on unphysical RSs close to the physical region. The situation of the poles for Model~II (shown in Tab.~\ref{1npII}) is similar to that of Model~I. The lower bound state mainly couples to the $D^*\bar{D}^*$ channel. Besides the second resonance, the other resonances mainly couple to the $D_s^{*+}D_s^{*-}$ channel. These $1^{-+}$ exotic states can be searched for in the electron-positron annihilation process with an emission of one photon~\cite{Du:2016qcr,Zhang:2025gmm}. 

\section{SUMMARY and outlook}
\label{sec:summary}
We perform a phenomenological study on the cross sections of the 
$e^+e^-\to D\bar{D}$, $e^+e^-\to D\bar{D}^*+c.c.$, $e^+e^-\to D^*\bar{D}^*$ processes. 
By constructing $P$-wave contact interaction between the $s_l^P=\frac{1}{2}^-$ HQSS doublet $(D,D^*)$ and its antiparticle,
we do a global analysis for the energy region $[3.7,4.25]~\mathrm{GeV}$, especially focusing on the property of the newly observed $G(3900)$. The upper limit energy is 
restricted by the next opening threshold $D_1\bar{D}$. To accommodate the open-charmed-strange meson pair channels, we work within the SU(3) flavor symmetry framework.
In the considered energy region, there are three well-established charmonia, i.e. $\psi(1D)$, $\psi(3S)$ and $\psi(2D)$, which affect the cross sections. We work in two models for a comparison: three bare charmonia scenario (Model~I) and two bare charmonia scenario (Model~II). In Model~I (Model~II), we find three (two) renormalized poles corresponding to the input bare poles. Besides these poles, we find a pole at $3832.57^{+0.91}_{-0.79}\pm 74.53^{+0.68}_{-2.15}i~\mathrm{MeV}$, about 40~MeV below the $[D\bar{D}^*]_{\text{Thr}}$, on the $(-,+,-,+,+,+)$ sheet in Model~I. In Model~II, a pole $3883.91^{+0.38}_{-0.46}\pm 46.53^{+1.22}_{-1.22}i$~MeV on the $(-,-,+,+,+,+)$ sheet is 9~MeV above the $[D\bar{D}^*]_{\text{Thr}}$ threshold, connecting to the physical sheet above the $D\bar{D}^*$ threshold and below the $D_s^+D_s^-$ threshold. Both of them are dynamically generated states based on the trajectory of the pole  renormalization. In this sense, we conclude that the $G(3900)$ is a dynamically generated state. With the parameters fixed in the $J^{PC}=1^{--}$ channel, we also predict several dynamically generated states in the $J^{PC}=1^{-+}$ channel, which can be investigated in the electron-positron annihilation process involving the emission of a single photon.

\bigskip

\section*{Acknowledgments}
We are grateful to Christoph Hanhart for fruitful discussions.
This work is partly supported by the National Natural Science Foundation of China with Grant Nos.~12375073 and ~12035007, Guangdong Provincial funding with Grant No.~2019QN01X172. 
The work of UGM was supported in part  by the CAS President's International Fellowship Initiative (PIFI) under Grant No.~2025PD0022
and by the MKW NRW under the funding code No.~NW21-024-A.

\nocite{*}
\bibliography{ref.bib}

\onecolumngrid
\newpage
\appendix

\clearpage
{\large \bf \section*{Supplemental Materials}}

\section{The production amplitude}\label{pro_amp}
According to Ref.~\cite{Zou:2002ar}, the covariant amplitude is built from pure orbital angular momentum covariant tensors and covariant spin wave functions $\phi_{\mu_1\dots \mu_s}$ which are interpreted as the polarization vectors of the final state particles, together with the operators $g_{\mu\nu}$, $\epsilon_{\mu\nu\lambda\sigma}$ and momenta of parent particles. From Ref.~\cite{Chung:1993da}, we can find that the explicit expression for the pure orbital angular momentum covariant tensors with relative orbital angular momentum $l=1$ in $a\to b+c$ process reads as
\begin{align}
    t_{\mu}^{(1)}= \widetilde{g}_{\mu\nu}(p_a)r^{\nu}
    \label{orb}
\end{align}
where 
\begin{align}
    \widetilde{g}^{\mu\nu}(p_a)=g^{\mu\nu}-\frac{p_a^{\mu} p_a^{\nu}}{p_a^2},\quad r=p_b-p_c.
    \label{orb_1}
\end{align}

The projection operator for spin-0 and spin-2 are
\begin{align}
    P^{(0)}_{\alpha\beta} & =\frac{1}{\sqrt{3}}g_{\alpha\beta},\notag\\
    P^{(2)}_{\alpha\beta\gamma\delta} & = \frac{1}{2}(\widetilde{g}_{\alpha\gamma}\widetilde{g}_{\beta\delta}+\widetilde{g}_{\alpha\delta}\widetilde{g}_{\beta\gamma})-\frac{1}{3}\widetilde{g}_{\alpha\beta}\widetilde{g}_{\gamma\delta},
    \label{spin_1}
\end{align}
respectively.

In the following, we show how Eqs.~(\ref{Amp_1})–(\ref{Amp_4}) are obtained by combining Eqs. (\ref{orb})–(\ref{spin_1}). Eq.~(\ref{orb}) provides the covariant amplitude for the process $a\to b+c$ with $l=1$. However, in our production vertex, the particle a corresponds to a virtual photon rather than a real particle. As a result, the factor $\widetilde{g}_{\mu\nu}$ should be absorbed into the photon propagator, as shown in Eq.~(\ref{sca_amp}). Since the photon has a transversal polarization, only the transverse part contributes. Therefore, Eq.~(\ref{spin_1}) should be rewritten as
\begin{align}
    P^{(2)}_{ij,mn} & = \frac{1}{2}(g_{im}g_{jn}+g_{in}g_{jm})-\frac{1}{3}g_{ij}g_{mn}.
    \label{spin_2}
\end{align}
Due to Eq. (\ref{Amp}), we have reduced the Lorentz indices from four-dimensional to three-dimensional form. As a result, all the indices in Eq. (\ref{spin_2}) refer to three-dimensional components. The production amplitude is required to contain a Lorentz index so that it can be contracted with the photon propagator. 

\begin{enumerate}
    \item For $\gamma^*\to D\bar{D}$ process, $l=1$ ,$S=0$ and $J^{PC}=1^{--}$. Since the polarization vector $\varepsilon_{D}=1$, the production amplitude takes the form:
    \begin{align}
        \mathcal{A}_1^i= \mathcal{U}_1 r^i = \mathcal{U}_1 (p_{\bar{D}}-p_{D})^i.
    \end{align}
    Here $\mathcal{U}_i$ is the physical production amplitude, which can be interpreted as a form factors. 
    \item For $\gamma^*\to D\bar{D}^*+c.c.$ process, $l=1$, $S=1$ and $J^{PC}=1^{--}$. Since the $D^*$ meson provides a polarization vector $\varepsilon_{\lambda k}^*$ with one Lorentz index and the relative momentum also carries a Lorentz index, we introduce the tensor $\epsilon^{ijk}$  perform index contraction, so that the production amplitude contains only a single Lorentz index. The production amplitude reads as 
    \begin{align}
        \mathcal{A}_2^i= \mathcal{U}_2\epsilon^{ijk} r_j \varepsilon_{\lambda k}^* = \mathcal{U}_2 \epsilon^{ijk}(p_{\bar{D}}-p_{D})_j \varepsilon_{\lambda k}^*,
    \end{align}
    where $\lambda$ is the helicity index.
    \item For $\gamma^*\to D^*\bar{D}^*_{S=0}$ process, $l=1$, $S=0$ and $J^{PC}=1^{--}$. In this case, the polarization vectors of $D^*$ and $\bar{D}^*$ mesons form a rank-2 tensor $\varepsilon_{\lambda}^{*m}\varepsilon_{\lambda'}^{*n}$. The coupling of two vector particles allows total spin 0, 1 or 2. Since we focus on spin-0 case, we employ the spin-0 projection operator to extract the corresponding spin 0. The production amplitude reads as
    \begin{align}
        \mathcal{A}_3^i & = \mathcal{U}_3 P^{(0)~mn} r_j \varepsilon_{\lambda~m}^{*}\varepsilon_{\lambda'~n}^{*},\notag\\
        & = \frac{1}{\sqrt{3}}\mathcal{U}_3(p_{\bar{D}^{*}}-p_{D^*})^i\varepsilon_{\lambda}^{*}\cdot\varepsilon_{\lambda'}^{*}.
    \end{align}
    \item For $\gamma^*\to D^*\bar{D}^*_{S=2}$ process, $l=1$, $S=2$ and $J^{PC}=1^{--}$. Similarly, we employ the spin-2 projection operator to extract the corresponding spin 2. The production amplitude reads as 
    \begin{align}
        \mathcal{A}_4^i & = \mathcal{U}_4 P^{(2)~ij,mn} r_j \varepsilon_{\lambda~m}^{*}\varepsilon_{\lambda'~n}^{*},\notag\\
        & \to \sqrt{\frac{3}{5}} \mathcal{U}_4 P^{(2)~ij,mn} (p_{\bar{D}^{*}}-p_{D^*})_j \varepsilon_{\lambda~m}^{*}\varepsilon_{\lambda'~n}^{*}.
    \end{align}
    In the last step, the normalization factor $\sqrt{3/5}$ is multiplied.
\end{enumerate}
Only when the normalization factor is considered in the amplitude level, all the four amplitude squares satisfy $A_n^iA_n^i=\mathcal{U}_n|p_{\bar{D}^{(*)}}-p_{D^{(*)}}|^2$ for $n=1,2,3,4$.

\section{The detailed calculation on amplitude squared}\label{detail cal}
The amplitudes squared for $e^+e^- \rightarrow (D^{(*)}\bar{D}^{(*)})^a_n$ read as
\begin{align}
    |\overline{\mathcal{M}_n^a}|^2 =\frac{1}{2}\sum_r\frac{1}{2}\sum_s\sum_{\lambda}\sum_{\lambda'}|\mathcal{M}^a_n|^2 & = \frac{e^2}{4s^2}\sum_r\sum_s\sum_{\lambda}\sum_{\lambda'}\bar{v}^r(p_+)\gamma^{\nu}u^s(p_-)\bar{u}^s(p_-)\gamma^{\nu'}v^r(p_+)\mathcal{A}^a_{n\nu}\mathcal{A}^{*a}_{n\nu'}\notag\\
    & = \frac{e^2}{4s^2}\sum_{\lambda}\sum_{\lambda'}\mathrm{Tr}[\slashed{p}_+\gamma^\nu\slashed{p}_-\gamma^{\nu'}]\mathcal{A}^a_{n\nu}\mathcal{A}^{*a}_{n\nu'}\notag\\
    & = \frac{e^2}{s^2}\sum_\lambda\sum_{\lambda'}(p_+^{\nu}p_-^{\nu'}+p_+^{\nu'}p_-^{\nu}-g^{\nu\nu'}p_+p_-)\mathcal{A}^a_{n\nu}\mathcal{A}^{*a}_{n\nu'}.
\end{align}
In the center of mass frame, $s=4E^2$ and $p_+\cdot p_-=2 E^2$, where $E$ is the energy of electron. Plugging them into above equation, one can obtain
\begin{align}
    |\overline{\mathcal{M}_n^a}|^2  = & \frac{4\pi\alpha}{s^2}\sum_\lambda\sum_{\lambda'}(p_+^{\nu}p_-^{\nu'}+p_+^{\nu'}p_-^{\nu}-g^{\nu\nu'}p_+p_-)\mathcal{A}^a_{n\nu}\mathcal{A}^{*a}_{n\nu'} \notag\\
     = & \sum_\lambda\sum_{\lambda'}\left[\frac{4\pi\alpha}{s^2}(p_+^0p_-^0+p_+^0p_-^0)\mathcal{A}^a_{n0}\mathcal{A}^{*a}_{n0}+\frac{4\pi\alpha}{s^2}(p_+^0p_-^j+p_+^jp_-^0)\mathcal{A}^a_{n0}\mathcal{A}^{*a}_{nj}\right.\notag\\
    & +\frac{4\pi\alpha}{s^2}(p_+^ip_-^0+p_+^0p_-^i)\mathcal{A}^a_{ni}\mathcal{A}^{*a}_{n0} + \frac{4\pi\alpha}{s^2}(p_+^ip_-^j+p_+^jp_-^i)\mathcal{A}^a_{ni}\mathcal{A}^{*a}_{nj}\notag\\
    & \left .-\frac{4\pi\alpha}{s^2}\frac{1}{2}g^{00}s\mathcal{A}_{n0}^a\mathcal{A}^{*a}_{n0}-\frac{4\pi\alpha}{s^2}\frac{1}{2}g^{ij}s\mathcal{A}^a_{ni}\mathcal{A}^{*a}_{nj}\right]\notag\\
     = &
    -\frac{4\pi\alpha}{s^2}\sum_{\lambda}\sum_{\lambda'}(\frac{1}{2}sg^{ij}+2p_+^ip_+^j)\mathcal{A}^a_{ni}\mathcal{A}^{*a}_{nj} \quad\quad (i,j = 1,2,3).
\end{align}
Substituting Eqs.(\ref{Amp_1})$-$(\ref{Amp_4}) into above equation and using relation $p_+^i\cdot (p_{\bar{D}^{(*)}}-p_{D^{(*)}})_i=-2E|p_{D^{(*)}}|cos\theta$ and $(p_{\bar{D}^{(*)}}-p_{D^{(*)}})_i(p_{\bar{D}^{(*)}}-p_{D^{(*)}})^i=-4|p_{D^{(*)}}|^2$ in the center of mass frame, one can obtain
\begin{align}
    |\overline{\mathcal{M}_1^a}|^2 & =-\frac{4\pi\alpha}{s^2}\sum_{\lambda}\sum_{\lambda'}(\frac{1}{2}sg^{ij}+2p_+^ip_+^j)\mathcal{A}^a_{1i}\mathcal{A}_{1j}^{*a}\notag\\
    & = -\frac{4\pi\alpha}{s^2}(\frac{1}{2}sg^{ij}+2p_+^ip_+^j)\left[\mathcal{U}^a_1(p_{\bar{D}}-p_D)_i\right]\left[\mathcal{U}_1^{*a}(p_{\bar{D}}-p_D)_j\right]\notag\\
    & = -\frac{2\pi\alpha}{s}|\mathcal{U}^a_1|^2(p_{\bar{D}}-p_D)_i(p_{\bar{D}}-p_D)^i-\frac{8\pi\alpha}{s^2}|\mathcal{U}^a_1|^2\left[p_+^i\cdot (p_{\bar{D}}-p_D)_i\right][p_+^j\cdot (p_{\bar{D}}-p_D)_j]\notag\\
    & = \frac{8\pi\alpha}{s}|p_D|^2|\mathcal{U}_1^a|^2(1-\mathrm{cos}^2\theta),
\end{align}
\begin{align}
    |\overline{\mathcal{M}_2^a}|^2 & =-\frac{4\pi\alpha}{s^2}\sum_{\lambda}\sum_{\lambda'}(\frac{1}{2}sg^{ij}+2p_+^ip_+^j)\mathcal{A}^a_{2i}\mathcal{A}_{2j}^{*a}\notag\\
    & = -\frac{4\pi\alpha}{s^2}\sum_\lambda(\frac{1}{2}sg^{ij}+2p_+^ip_+^j)\left[\mathcal{U}^a_2\epsilon_{i\gamma k}(p_{\bar{D}}-p_{D^*})^\gamma\varepsilon^{*k}_\lambda\right]\left[\mathcal{U}_2^{*a}\epsilon_{j\alpha\beta}(p_{\bar{D}}-p_{D^*})^\alpha\varepsilon^{\beta}_\lambda\right]\notag\\
    & = -\frac{2\pi\alpha}{s}|\mathcal{U}^a_2|^2\epsilon_{i\gamma k}\epsilon^{i \alpha \beta}(p_{\bar{D}}-p_{D^*})^{\gamma}(p_{\bar{D}}-p_{D^*})_{\alpha}(-g^k_{\beta})-\frac{8\pi\alpha}{s^2}|\mathcal{U}^a_2|^2p_{+i}p_{+}^{j}\epsilon^{i\alpha\beta}\epsilon_{j\rho\sigma}(p_{\bar{D}}-p_{D^*})_{\alpha}(p_{\bar{D}}-p_{D^*})^{\rho}(-g^{\sigma}_{\beta})\notag\\
    & = \frac{2\pi\alpha}{s}|\mathcal{U}^a_2|^2(-2g_\gamma^\alpha)(p_{\bar{D}}-p_{D^*})^{\gamma}(p_{\bar{D}}-p_{D^*})_{\alpha}+\frac{8\pi\alpha}{s^2}|\mathcal{U}^a_2|^2(-g^i_jg^\alpha_\rho+g^i_\rho g^\alpha_j)p_{+i}(p_{\bar{D}}-p_{D^*})^{\rho}p_{+}^{j}(p_{\bar{D}}-p_{D^*})_{\alpha}\notag\\
    & = \frac{8\pi\alpha}{s}|\mathcal{U}^a_2|^2|p_{D}|^2(1+\mathrm{cos}^2\theta),
\end{align}
where the completeness relation $\sum_{\lambda=0,\pm 1}\varepsilon^*_{\lambda i}\varepsilon_{\lambda j}=-g_{ij}+\frac{p_ip_j}{m^2}$has been used, and it is easily to prove that only the first term will contribute to the amplitude squared.
Similarly,
\begin{align}
    |\overline{\mathcal{M}^a_3}|^2  = &-\frac{4\pi\alpha}{s^2}\sum_{\lambda}\sum_{\lambda'}(\frac{1}{2}sg^{ij}+2p_+^ip_+^j)\mathcal{A}^a_{3i}\mathcal{A}_{3j}^{*a}\notag\\
     = & -\frac{2\pi\alpha}{3s}\sum_{\lambda}\sum_{\lambda'}[(p_{\bar{D}^*}-p_{D^*})_i\varepsilon^{*\alpha}_\lambda\varepsilon^*_{\lambda'\alpha}][(p_{\bar{D}^*}-p_{D^*})^i\varepsilon^{\rho}_{\lambda}\varepsilon_{{\lambda'}\rho}]|\mathcal{U}^a_3|^2\notag\\
     & -\frac{8\pi\alpha}{3s^2}\sum_{\lambda}\sum_{\lambda'}p_+^ip_+^j[(p_{\bar{D}^*}-p_{D^*})_i\varepsilon^{*\alpha}_\lambda\varepsilon^*_{\lambda'\alpha}][(p_{\bar{D}^*}-p_{D^*})_j\varepsilon^{\rho}_{\lambda}\varepsilon_{{\lambda'}\rho}]|\mathcal{U}^a_3|^2\notag\\
     = & -\frac{2\pi\alpha}{3s}(p_{\bar{D^*}}-p_{D^*})_i(p_{\bar{D^*}}-p_{D^*})^i(-g^{\alpha\rho})(-g_{\alpha\rho})|\mathcal{U}^a_3|^2-\frac{8\pi\alpha}{3s^2}p^i_+p^j_+(p_{\bar{D^*}}-p_{D^*})_i(p_{\bar{D^*}}-p_{D^*})_j(-g^{\alpha\rho})(-g_{\alpha\rho})|\mathcal{U}^a_3|^2\notag\\
     = & \frac{8\pi\alpha}{s}|p_{D^*}|^2|\mathcal{U}^a_3|^2(1-\mathrm{cos}^2\theta),
\end{align}
\begin{align}
    |\overline{\mathcal{M}^a_4}|^2  = &-\frac{4\pi\alpha}{s^2}\sum_{\lambda}\sum_{\lambda'}(\frac{1}{2}sg^{ij}+2p_+^ip_+^j)\mathcal{A}^a_{4i}\mathcal{A}_{4j}^{*a}\notag\\
     = & -\frac{2\pi\alpha}{s}\sum_{\lambda}\sum_{\lambda'}[P_{ij,mn}(p_{\bar{D}^*}-p_{D^*})^j\varepsilon^{*m}_\lambda\varepsilon^{*n}_{\lambda'}][P^{i\alpha,\beta\gamma}(p_{\bar{D}^*}-p_{D^*})_\alpha\varepsilon_{\lambda\beta}\varepsilon_{\lambda'\gamma}]|\mathcal{U}^a_4|^2\notag\\
     & -\frac{8\pi\alpha}{s^2}\sum_{\lambda}\sum_{\lambda'}p_+^ip_+^j[P_{i\alpha,mn}(p_{\bar{D}^*}-p_{D^*})^\alpha\varepsilon^{*m}_\lambda\varepsilon^{*n}_{\lambda'}][P_{j\beta,\rho\sigma}(p_{\bar{D}^*}-p_{D^*})^\beta\varepsilon_{\lambda}^{\rho}\varepsilon_{\lambda'}^{\sigma}]|\mathcal{U}^a_4|^2\notag\\
     = & -\frac{2\pi\alpha}{s}P_{ij,mn}P^{i\alpha,\beta\gamma}(p_{\bar{D}^*}-p_{D^*})^j(p_{\bar{D}^*}-p_{D^*})_\alpha|\mathcal{U}^a_4|^2(-g^m_\beta)(-g^n_\gamma)\notag\\
     & -\frac{8\pi\alpha}{s^2}p_+^ip_+^jP_{i\alpha,mn}P_{j\beta,\rho\sigma}(p_{\bar{D}^*}-p_{D^*})^{\alpha}(p_{\bar{D}^*}-p_{D^*})^{\beta}|\mathcal{U}^a_4|^2(-g^{m\rho})(-g^{n\sigma})\notag\\
     = & -\frac{2\pi\alpha}{s}P_{ij,mn}P^{i\alpha,mn}(p_{\bar{D}^*}-p_{D^*})^j(p_{\bar{D}^*}-p_{D^*})_\alpha|\mathcal{U}^a_4|^2-\frac{8\pi\alpha}{s^2}p_+^ip_+^jP_{i\alpha,mn}P_{j\beta}^{mn}(p_{\bar{D}^*}-p_{D^*})^\alpha(p_{\bar{D}^*}-p_{D^*})^{\beta}|\mathcal{U}^a_4|^2\notag\\
     = & -\frac{2\pi\alpha}{s}\delta_j^\alpha(p_{\bar{D}^*}-p_{D^*})^j(p_{\bar{D}^*}-p_{D^*})_\alpha|\mathcal{U}^a_4|^2\notag\\
     & -\frac{8\pi\alpha}{s^2}(\frac{3}{10}\delta_{ij}\delta_{\alpha\beta}+\frac{3}{10}\delta_{i\beta}\delta_{i\alpha}-\frac{1}{5}\delta_{i\alpha}\delta_{j\beta})p_+^ip_+^j(p_{\bar{D}^*}-p_{D^*})^\alpha(p_{\bar{D}^*}-p_{D^*})^\beta|\mathcal{U}^a_4|^2\notag\\
     = & \frac{28\pi\alpha}{5s}|p_{D^*}|^2|\mathcal{U}^a_4|^2(1-\frac{1}{7}\mathrm{cos}^2\theta),
\end{align}
where we have used the relation
\begin{align}
    P_{ij,mn}P^{i\alpha,mn} & =\frac{3}{5}(\frac{1}{2}\delta_{im}\delta_{jn}+\frac{1}{2}\delta_{in}\delta_{jm}-\frac{1}{3}\delta_{ij}\delta_{mn})(\frac{1}{2}\delta^{im}\delta^{\alpha n}+\frac{1}{2}\delta^{in}\delta^{\alpha m}-\frac{1}{3}\delta^{i\alpha}\delta^{mn})=\delta^\alpha_j,\notag\\
    P_{i\alpha,mn}P_{j\beta}^{mn} & =\frac{3}{5}(\frac{1}{2}\delta_{im}\delta_{\alpha n}+\frac{1}{2}\delta_{in}\delta_{\alpha m}-\frac{1}{3}\delta_{i\alpha}\delta_{mn})(\frac{1}{2}\delta_j^m\delta_{\beta}^n+\frac{1}{2}\delta_{j}^{n}\delta_{\beta}^m-\frac{1}{3}\delta_{j\beta}\delta^{mn})\notag\\
    & =\frac{3}{10}\delta_{ij}\delta_{\alpha\beta}+\frac{3}{10}\delta_{i\beta}\delta_{j\alpha}-\frac{1}{5}\delta_{i\alpha}\delta_{j\beta}.
\end{align}

\clearpage
\section{The $P$-wave two-point function in the non-relativistic limit}\label{Green_fun}
In the relativistic expression, the two-point function reads as
\begin{align}
    B(E,m_1,m_2) & =i\int_a^b\frac{d^4q}{(2\pi)^4}\frac{f(|\Vec{q}|^2)}{(q^2-m_1^2+i\varepsilon^+)((p-q)^2-m_2^2+i\varepsilon^+)}\notag\\
    & =i\int_a^b\frac{d^4q}{(2\pi)^4}\frac{f(|\Vec{q}|^2)}{q_0^2-(|\Vec{q}|^2+m_1^2)+i\varepsilon^+)((E-q_0)^2-(|\Vec{q}|^2+m_2^2)+i\varepsilon^+)}\notag\\
    & = i\int_a^b\frac{d^4q}{(2\pi)^4}\frac{f(|\Vec{q}|^2)}{(q_0^2-\omega_1^2+i\varepsilon^+)((E-q_0)^2-\omega_2^2+i\varepsilon^+)},
    \label{NR_fun}
\end{align}
where $E$ is the center-of-mass energy and $\omega_i=\sqrt{|\Vec{q}|^2+m_i^2}$ with $i=1,2$. Here, $f(|\Vec{q}|^2)$ is a form factor, whose specific form depends on the truncation scheme. In the non-relativistic approximation, $\Vec{q}\rightarrow 0$, one can rewrite the denominator of above equation
\begin{align}
    q_0^2-\omega_1^2+i\varepsilon^+ & =(q_0+\omega_1-i\varepsilon)(q_0-\omega_1+i\epsilon)\notag\\
    & \approx (q_0+m_1+\frac{|\Vec{q}|^2}{2m_1}-i\varepsilon^+)(q_0-m_1-\frac{|\Vec{q}|^2}{2m_1}+i\varepsilon^+)\notag\\
    & \approx 2m_1(q_0-m_1-\frac{|\Vec{q}|^2}{2m_1}+i\varepsilon^+).
\end{align}
Similarly
\begin{align}
    (E-q_0)^2-\omega_2^2+i\varepsilon^+ & =(E-q_0+\omega_2-i\varepsilon^+)(E-q_0-\omega_2+i\varepsilon^+)\notag\\
    & \approx (E-q_0+m_2+\frac{|\Vec{q}|^2}{2m_2}-i\varepsilon^+)(E-q_0-m_2-\frac{|\Vec{q}|^2}{2m_2}+i\varepsilon^+)\notag\\
    & \approx 2m_2(E-q_0-m_2-\frac{|\Vec{q}|^2}{2m_2}+i\varepsilon^+).
\end{align}
Therefore, Eq.(\ref{NR_fun}) can be rewritten as
\begin{align}
     B(E,m_1,m_2) & =\frac{i}{4m_1m_2}\int_a^b\frac{d^4q}{(2\pi)^4}\frac{f(|\Vec{q}|^2)}{(q_0-m_1-\frac{|\Vec{q}|^2}{2m_1}+i\varepsilon^+)(E-q_0-m_2-\frac{|\Vec{q}|^2}{2m_2}+i\varepsilon^+)}\notag\\
     & = \frac{1}{4m_1m_2}\int\frac{d^3q}{(2\pi)^3}\frac{f(|\Vec{q}|^2)}{E-m_1-m_2-\frac{|\Vec{q}|^2}{2\mu}+i\varepsilon^+}\notag\\
     & = \frac{2\mu}{4m_1m_2}\int\frac{d^3q}{(2\pi)^3}\frac{f(|\Vec{q}|^2)}{k^2-|\Vec{q}|^2+i\varepsilon^+},
\end{align}
with $k=\sqrt{2\mu(E-m_1-m_2)}$. In the calculations of this paper, we have neglected the $1/(4m_1m_2)$ factor. Since the factor can be obtained by dividing $\prod_{i}\sqrt{2m_i}$ with $m_i$ the masses of the particle fields in the corresponding vertex, it can be absorbed by the fitting parameters of the contact interaction, and these parameters will add a squared energy dimension.

In order to calculate the $P$-wave two-point function, we need to compute the $S$-wave two-point function firstly
\begin{align}
    G_{S}(E) & =2\mu\int\frac{d^3q}{(2\pi^3)}\frac{e^{-\frac{2|\Vec{q}|^2}{\Lambda^2}}}{k^2-|\Vec{q}|^2+i\varepsilon^+}\notag\\
    & =-\frac{\mu\Lambda}{(2\pi)^{3/2}}+\frac{\mu k^2}{\pi^2}e^{-\frac{2|\Vec{q}|^2}{\Lambda^2}}\mathcal{P}\int_{0}^{\infty}\frac{d|\Vec{q}|}{k^2-|\Vec{q}|^2}e^{-\frac{2(k^2-|\Vec{q}|^2)}{\Lambda^2}}-i\frac{\mu k}{2\pi}e^{-\frac{2|\Vec{q}|^2}{\Lambda^2}},
    \label{G_s}
\end{align}
where we use a Gaussian form factor to regulate the ultraviolet divergence. 
Here, the Cauchy principal value integral is used to simplify above equation
\begin{align}
    \frac{1}{k^2-|\Vec{q}|^2+i\varepsilon^+} = \mathcal{P}\frac{1}{k^2-|\Vec{q}|^2}-i\frac{\pi}{2|\Vec{q}|}\delta(k-|\Vec{q}|).
\end{align}
One can define the function
\begin{align}
    f(x)=\mathcal{P}\int_{0}^{\infty}\frac{1}{k^2-|\Vec{q}|^2}e^{x(k^2-|\Vec{q}|^2)},
\end{align}
whose first derivative reads as
\begin{align}
    f'(x) & =\mathcal{P}\int_{0}^{\infty}d|\Vec{q}|e^{x(k^2-|\Vec{q}|^2)}=\frac{\sqrt{\pi}}{2\sqrt{x}}e^{(\sqrt{x}k)^2}.
\end{align}
According to the Newton-Leibniz formula, one can obtain
\begin{align}
    f(x)-f(0) & =\frac{\sqrt{\pi}}{2}\int_0^x dt \frac{1}{\sqrt{t}}e^{(\sqrt{t}k)^2}\notag\\
    & = \frac{\sqrt{\pi}}{k}\int_0^{\sqrt{x}k}d(\sqrt{t}k)e^{(\sqrt{t}k)^2}\notag\\
    & = \frac{\pi}{k}\mathrm{erfi}(\sqrt{x}k),
\end{align}
where $\text{erfi}(z)=\frac{2}{\sqrt{\pi}}\int_0^z dt e^{t^2}$ is the imaginary error function. It is easy to obtain $f(0)=0$, therefore, one can obtain
\begin{align}
    f(\frac{2}{\Lambda^2})=\mathcal{P}\int_{0}^{\infty}\frac{1}{k^2-|\Vec{q}|^2}e^{\frac{2}{\Lambda^2}(k^2-|\Vec{q}|^2)}=\frac{\pi}{2k}\mathrm{erfi}(\frac{\sqrt{2}}{\Lambda}k)
    \label{F_lambda}
\end{align}
Substituting Eq.(\ref{F_lambda}) into Eq.(\ref{G_s}), one can obtain
\begin{align}
G_{S}(E) & =-\frac{\mu\Lambda}{(2\pi)^{3/2}}+\frac{\mu k}{2 \pi}e^{-\frac{2k^2}{\Lambda^2}}\left[\text{erfi}(\frac{\sqrt{2}k}{\Lambda})-i\right]
\end{align}
There exists the following relationship between the $P$-wave two-point function and the first derivative of the $S$-wave two-point function
\begin{align}
    G_{P}(E) & =2\mu\int\frac{d^3q}{(2\pi^3)}\frac{|\Vec{q}|^2e^{-\frac{2|\Vec{q}|^2}{\Lambda^2}}}{k^2-|\Vec{q}|^2+i\varepsilon}=\frac{\Lambda^3}{4}\frac{\partial G_S(E)}{\partial \Lambda}\notag\\
    & = \frac{\Lambda^3}{4}\frac{\partial}{\partial \Lambda}\left(-\frac{\mu\Lambda}{(2\pi)^{3/2}}+\frac{\mu k}{2 \pi}e^{-\frac{2k^2}{\Lambda^2}}\left[\text{erfi}(\frac{\sqrt{2}k}{\Lambda})-i\right]\right)\notag\\
    & = -\frac{\mu \Lambda}{(2\pi)^{3/2}}(k^2+\frac{\Lambda^2}{4})+\frac{\mu k^3}{2\pi}e^{-\frac{2k^2}{\Lambda^2}}\left[\text{erfi}(\frac{\sqrt{2}k}{\Lambda})-i\right].
\end{align}
\clearpage
\section{Fitted parameters and effective couplings of poles }\label{FIT}
\begin{table}[hbt!]
\renewcommand{\arraystretch}{1.5}
    \centering
    \caption{The fitted parameters of Model I and Model II. The parameters $C_n^i$ and $g_{1D}^0$, $g_{3S}^0$ and $g_{2D}^0$ are contact interaction defined in Eqs.~(\ref{con_1})$-$(\ref{con_4}) and bare couplings between charmonium and charmed meson pair, respectively. $F_{S,D}^i$ and $f_{1D}^0$, $f_{3S}^0$ and $f_{2D}^0$ are the coupling of the virtual photon to the charmed meson pair and the charmonium, respectively. $m_{1D}$, $m_{3S}$ and $m_{2D}$ denote the bare masses of charmounia $\psi(1D)$, $\psi(3S)$ and $\psi(2D)$. }
    \label{tab2:parameter}
    \begin{tabular}{p{5.2cm}<{\centering}p{5.2cm}<{\centering}p{5.2cm}<{\centering}}
    \hline\hline
   \textbf{Parameters}  &  \textbf{Model I} & \textbf{Model II} \\
    \hline
   $C_1^0 ~[\mathrm{GeV^{-4}}]$  & $-672.91\pm 8.39$    & $-593.56\pm 17.11$  \\
   $C_2^0~[\mathrm{GeV^{-4}}]$  & $182.93\pm 15.36$  &  $-109.96 \pm 16.28$  \\
   $C_3^0~[\mathrm{GeV^{-4}}]$  & $-0.11\pm 10.60$  & $797.37\pm 32.52$  \\ 
   $C_4^0~[\mathrm{GeV^{-4}}]$  & $613.97\pm 17.17$ & $9.28\pm 9.4$  \\
   $C_1^8~[\mathrm{GeV^{-4}}]$  & $-208.49\pm 16.96$    & $-357.46\pm 15.66$  \\
   $C_2^8~[\mathrm{GeV^{-4}}]$  & $15.25\pm 9.82$  &  $-63.08 \pm 12.13$  \\
   $C_3^8~[\mathrm{GeV^{-4}}]$  & $-33.28\pm 9.43$  & $109.12 \pm 26.50$  \\ 
   $C_4^8~[\mathrm{GeV^{-4}}]$  & $638.27\pm 26.30$ & $475.96\pm 38.51$  \\
   $C_1^1~[\mathrm{GeV^{-4}}]$  & $-1159.87\pm 19.31$    & $-739.76\pm 23.82$  \\
   $C_2^1~[\mathrm{GeV^{-4}}]$  & $321.28\pm 17.50$  &  $263.68 \pm 21.22$  \\
   $C_3^1~[\mathrm{GeV^{-4}}]$  & $375.02\pm 25.13$  & $-292.25\pm 8.56$  \\ 
   $C_4^1~[\mathrm{GeV^{-4}}]$  & $438.66\pm 17.70$ & $-223.61\pm 8.68$  \\
   $g_{1D}^0~[\mathrm{GeV^{-1}}]$  & $0.66\pm 0.04$    & $-12.93\pm 0.26$  \\
   $g_{3S}^0~[\mathrm{GeV^{-1}}]$  & $-14.66\pm 0.37$  &  $-14.11 \pm 0.96$  \\
   $g_{2D}^0~[\mathrm{GeV^{-1}}]$  & $-17.09\pm 0.23$  & $-$  \\ 
   $f_S^0~[\mathrm{GeV^{0}}]$  & $-12.82\pm 0.34$  & $-4.92\pm 0.48$  \\ 
   $f_D^0~[\mathrm{GeV^{0}}]$  & $10.16\pm 0.28$ & $-4.62\pm 0.29$  \\
   $f_S^8~[\mathrm{GeV^{0}}]$  & $-16.72\pm 0.30$  & $-20.63\pm 0.76$  \\ 
   $f_D^8~[\mathrm{GeV^{0}}]$  & $8.75\pm 0.24$ & $7.3\pm 0.46$  \\
   $f_S^1~[\mathrm{GeV^{0}}]$  & $10.13\pm 0.21$  & $21.75\pm 0.45$  \\ 
   $f_D^1~[\mathrm{GeV^{0}}]$  & $-3.01\pm 0.11$ & $-5.38\pm 0.16$  \\
   $f_{1D}^0~[\mathrm{GeV^{3}}]$  & $-0.30\pm 0.02$  & $0.13\pm 0.00$  \\ 
   $f_{3S}^0~[\mathrm{GeV^{3}}]$  & $-11.96\pm 0.63$ & $-0.47\pm 0.05$  \\
   $f_{2D}^0~[\mathrm{GeV^{3}}]$  & $-0.15\pm 0.00$ & $-$  \\
   $m_{1D}^0~[\mathrm{GeV}]$  & $3.807\pm 0.001$ & $3.804\pm 0.001$  \\
   $m_{3S}^0~[\mathrm{GeV}]$  & $4.229 \pm 0.002$ & $4.253\pm 0.005$ \\
   $m_{2D}^0~[\mathrm{GeV}]$  & $3.692\pm 0.003$  & $-$ \\
   $\Lambda~[\mathrm{GeV}]$ & $0.50 \pm 0.00$ & $0.50\pm 0.00$\\
   $\chi^2/\mathrm{d.o.f.}$ & 2.17  &  2.66\\
   \hline
   \hline
   \end{tabular}  
\end{table}

\begin{table}
\renewcommand{\arraystretch}{1.5}
    \centering
    \caption{Pole positions and effective couplings of $1^{--}$ system in Model I on various RSs. The dimension of coupling is $\mathrm{GeV}^{-3/2}$. The square brackets represent energy distance of the poles to the physical RS, with unit MeV.}
    \label{tab:1nnModelI}
    \begin{tabular}{p{2.5cm}<{\centering}p{3cm}<{\centering}p{1.1cm}<{\centering}p{1.1cm}<{\centering}p{1.1cm}<{\centering}p{1.3cm}<{\centering}p{1.3cm}<{\centering}p{1.3cm}<{\centering}p{1.5cm}<{\centering}p{1.5cm}<{\centering}}
    \hline\hline
     \textbf{RSs} &\textbf{Poles} [MeV] & $g_{D\bar{D}}$ &$g_{D\bar{D}^*}$ &  $g_{D_s^+D_s^-}$  & $ g_{{D^*\bar{D}_{s=0}^*}}$ & $ g_{{D^*\bar{D}_{s=2}^*}}$ & $g_{D_s^+D_s^{*-}}$ & $g_{{D_s^{*+}D_{s~s=0}^{*-}}}$ & $g_{{D_s^{*+}D_{s~s=2}^{*-}}}$ \\
    \hline
    $(+,+,+,+,+,+)$ & 3691.60 & 0.11 & 0.30 & 0.66& 0.30 & 0.24 & 2.10 & 3.18 & 2.01  \\
    $(-,+,+,+,+,+)$ & $3778.42\pm11.81i$ [12] & 1.31 & 2.72 & 8.54& 12.13 & 2.33 & 22.97 & 35.91 & 20.09  \\
    $(-,+,-,+,+,+)$ & $3832.52\pm74.53i$ & 1.02 & 4.29 & 0.14& 29.63 & 5.18 & 17.71 & 172.75 & 26.36  \\
    $(-,-,-,-,+,+)$ & $4011.05\pm 10.13i$ [16]& 0.16 & 0.32 & 0.34& 1.73 & 0.28 & 0.83 & 8.81 & 0.69  \\
    $(-,-,-,-,-,-)$ & $4232.78\pm 23.96i$ [24]& 0.02 & 0.08 & 0.12& 0.04 & 0.27 & 0.37 & 1.22 & 1.52  \\
   \hline
   \hline
    \\
    \\
   \end{tabular}  

    \centering
    \caption{Pole positions and effective couplings of $1^{--}$ system in Model II on various RSs. Other details are the same as Tab.~\ref{tab:1nnModelI}.}
    \label{tab:1nnModelII}
    \begin{tabular}{p{2.5cm}<{\centering}p{3cm}<{\centering}p{1.1cm}<{\centering}p{1.1cm}<{\centering}p{1.1cm}<{\centering}p{1.3cm}<{\centering}p{1.3cm}<{\centering}p{1.3cm}<{\centering}p{1.5cm}<{\centering}p{1.5cm}<{\centering}}
    \hline\hline
     \textbf{RSs} &\textbf{Poles $[\mathrm{MeV}]$} & $g_{D\bar{D}}$ &$g_{D\bar{D}^*}$ &  $g_{D_s^+D_s^-}$  & $ g_{{D^*\bar{D}_{s=0}^*}}$ & $ g_{{D^*\bar{D}_{s=2}^*}}$ & $g_{D_s^+D_s^{*-}}$ & $g_{{D_s^{*+}D_{s~s=0}^{*-}}}$ & $g_{{D_s^{*+}D_{s~s=2}^{*-}}}$ \\
    \hline
   $(-,+,+,+,+,+)$ & $3743.07\pm 7.36i$ [7] & 2.39 & 0.92 & 0.01& 19.15 & 6.28 & 0.02 & 0.10 & 0.03  \\
   $(-,+,+,+,+,+)$ & $3775.29\pm14.31i$ [14] & 1.55 & 4.24 & 8.94 &13.50  & 8.89 & 33.29 & 27.29 & 56.34  \\
    $(-,-,+,+,+,+)$ & $3883.91\pm46.53i$ [47]& 0.08 & 1.41 & 0.00& 2.57 & 8.68 & 0.01 & 0.03 & 0.01  \\
    $(-,-,-,-,+,+)$ & $4019.42\pm17.40i$ [17]& 0.21 & 0.24 & 0.22& 1.58 & 0.86 & 0.63 & 2.92 & 4.59  \\
    $(-,-,-,-,-,-)$ & $4278.21\pm 21.59i$ [22]& 0.03 & 0.09 & 0.12& 0.07 & 0.13 & 0.46 & 0.57 & 0.55  \\
   \hline
   \hline
   \end{tabular}  
\end{table}
\clearpage
\section{The trajectory of poles in Model I and the angular distribution of $D$ meson.}\label{all_tra}
\begin{figure}[hbt!]
    \centering
    \includegraphics[width=0.41\linewidth]{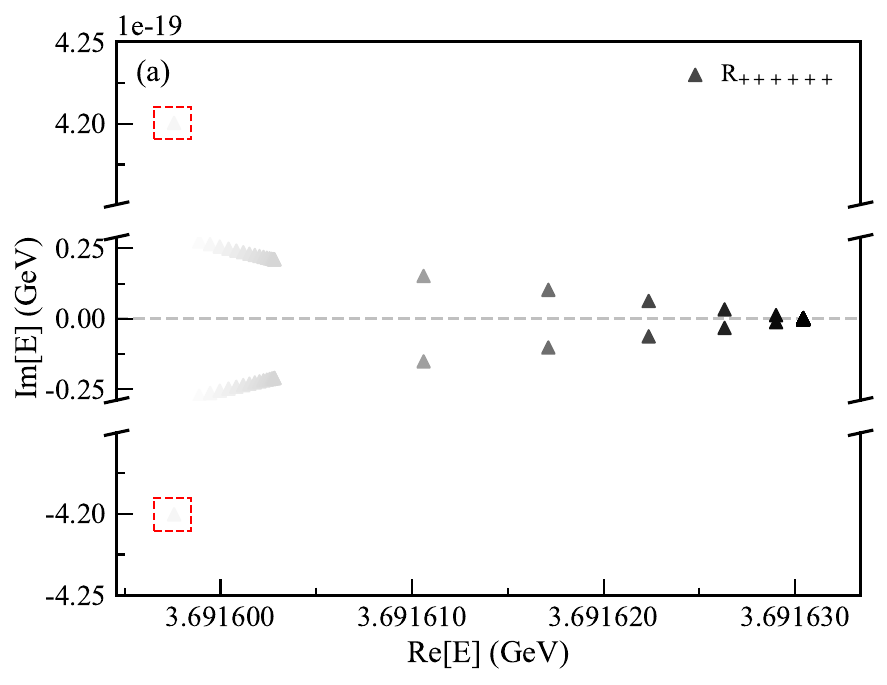}
    \includegraphics[width=0.41\linewidth]{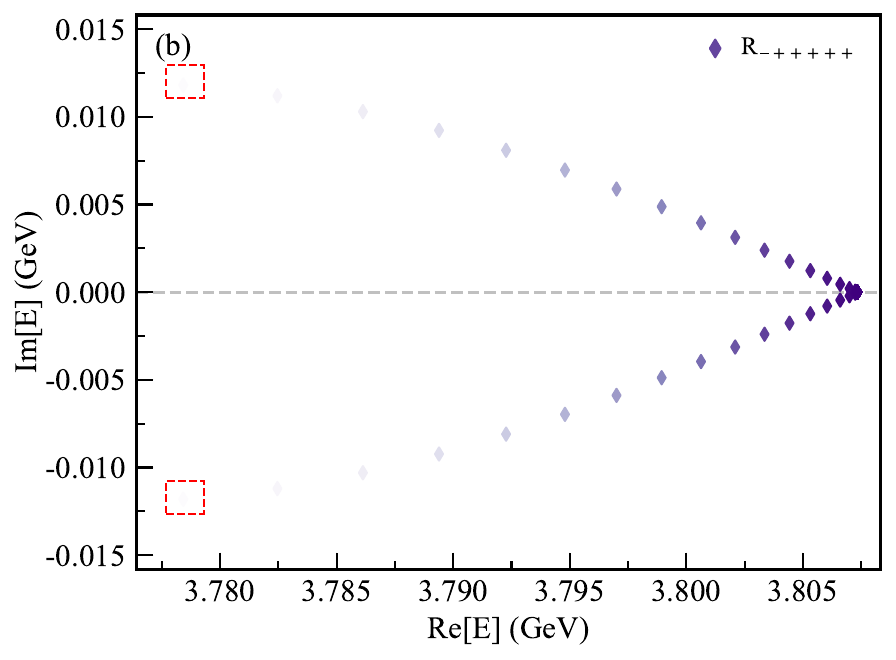}
    \includegraphics[width=0.41\linewidth]{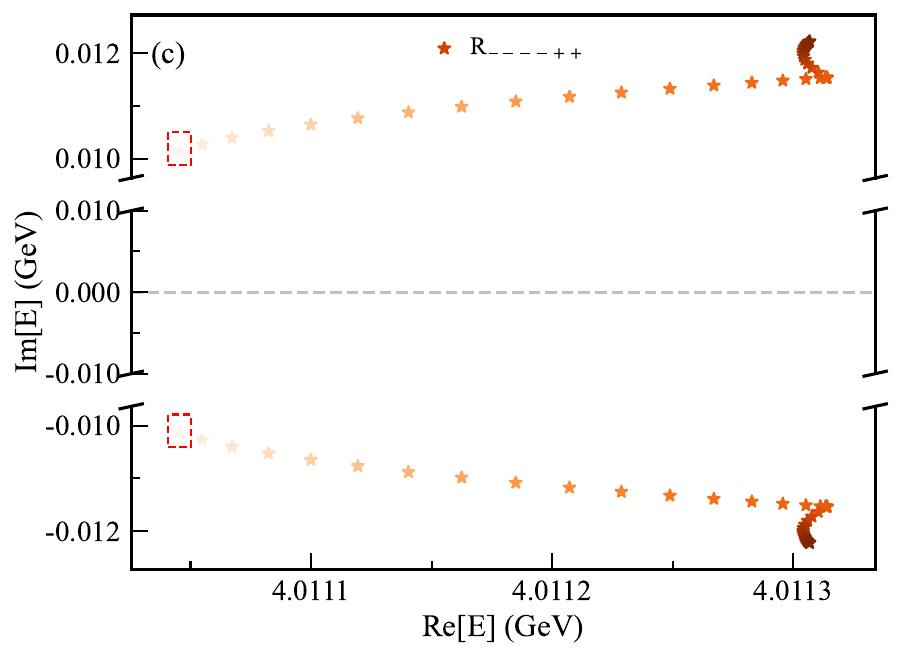}
    \includegraphics[width=0.41\linewidth]{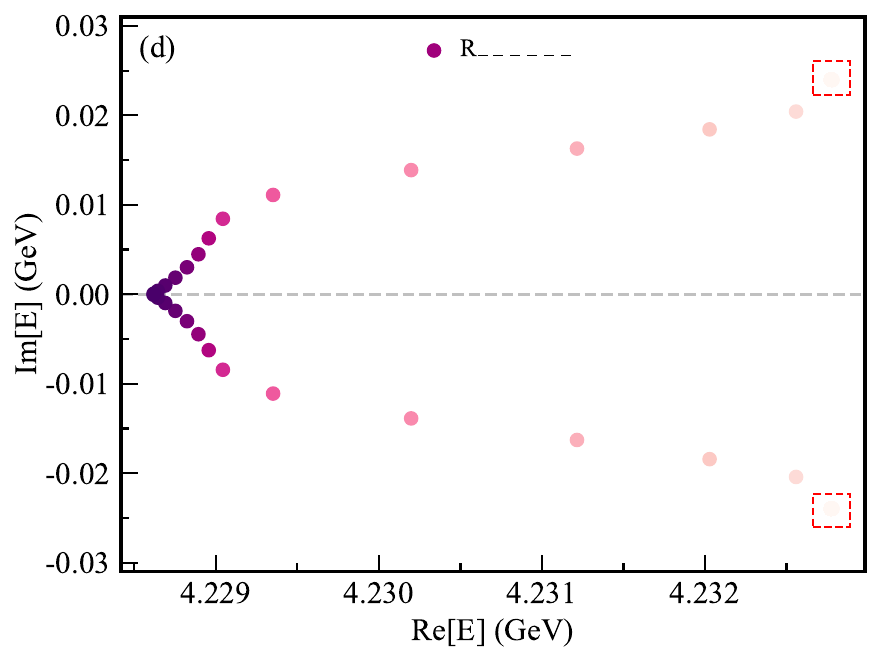}
    \caption{The zoomed-in trajectory of poles in Model I on various RSs with the couplings $g_{2D}^0$, $g_{1D}^0$ and $g_{3S}^0$ varying sequentially from the fitted values to zero. The $(a)$, $(b)$, $(c)$ and $(d)$ figures represent the trajectories of the poles 3691.60 GeV, $3778.42\pm 11.81i$ GeV, $4011.05\pm10.31i$ GeV, and $4232.78\pm 23.96i$ GeV. $R_{\pm,\dots,\pm}$ represent the RSs where the poles locate. The red dashed rectangular boxes represent the initial positions of the poles points represent the initial positions of the poles. The colors of the poles gradually become darker as the parameters $g_{2D}^0$, $g_{1D}^0$ and $g_{3S}^0$ vary sequentially from the fitted values to zero.} 
\end{figure}
\begin{figure}[hbt!]
    \centering
    \includegraphics[width=0.41\linewidth]{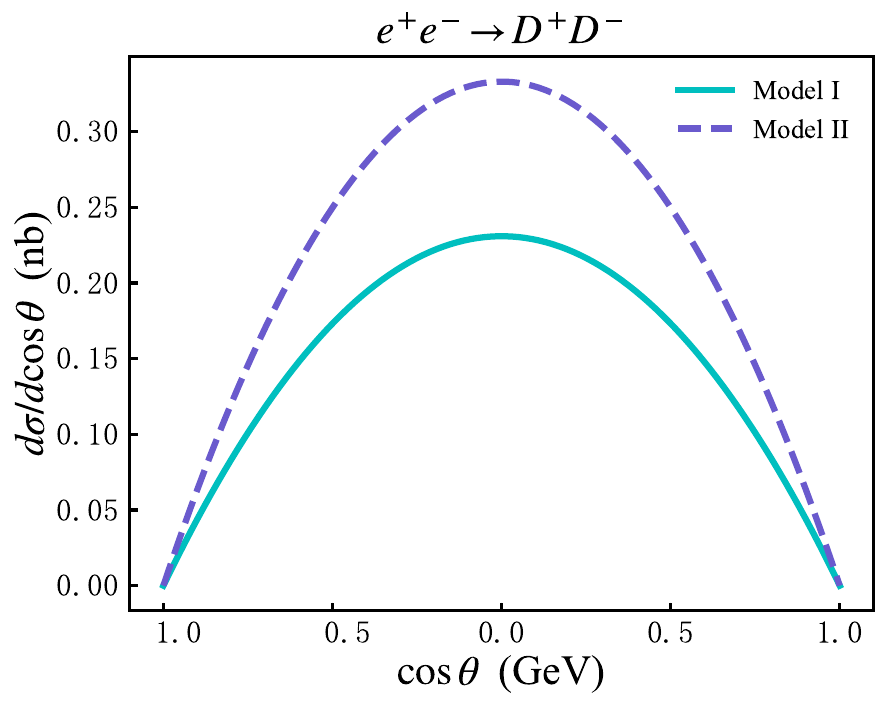}
    \includegraphics[width=0.41\linewidth]{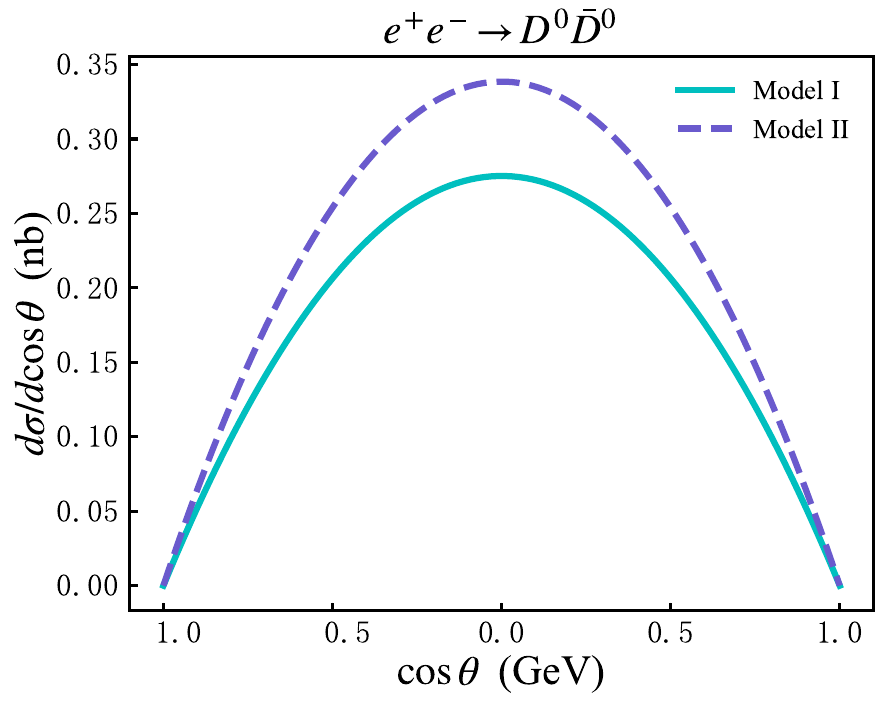}
    \caption{The $D$ meson scattering angle distribution at $\sqrt{s}=3.873$ GeV.}   
    \label{Dis}
\end{figure}

\section{The distribution of the standardized residuals for Model I and Model II}\label{residual}
\begin{figure}[hbt!]
    \centering
    \includegraphics[width=0.45\linewidth]{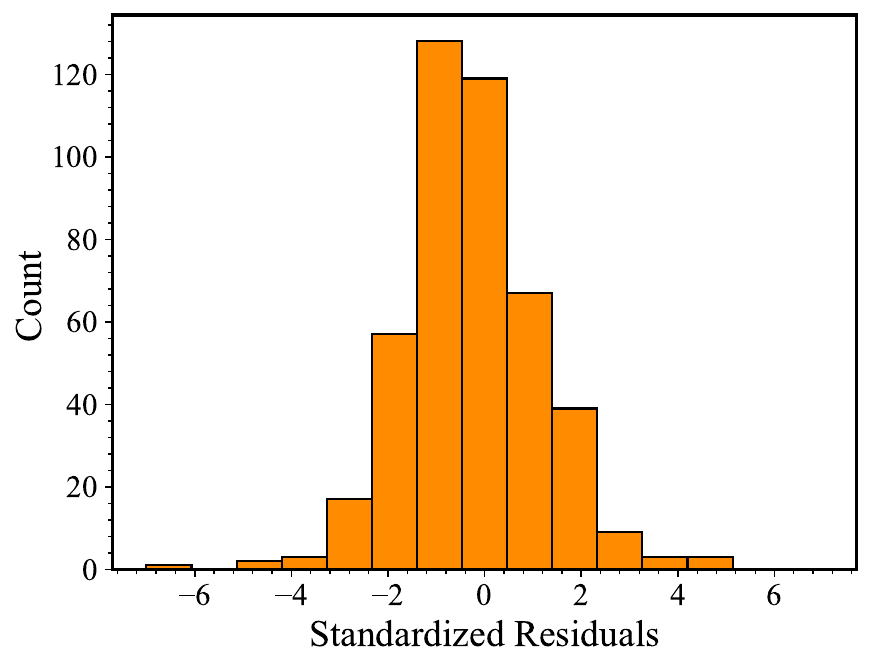}
    \includegraphics[width=0.45\linewidth]{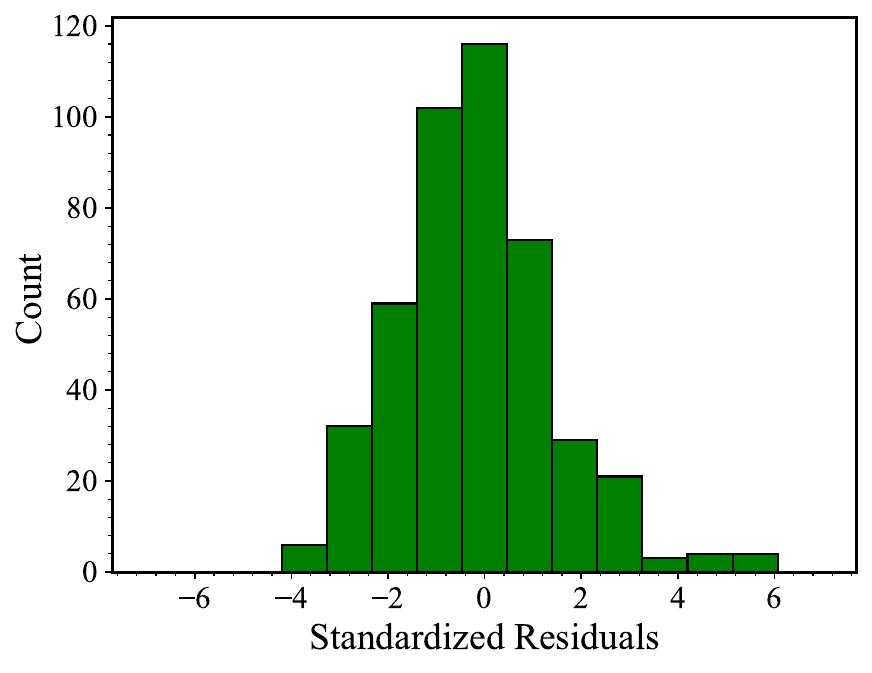}
    \caption{ The distribution of the standardized residuals for Models I (orange) and II (green). There are 15 bins in region [-7, 7].}   
    \label{SR15}
\end{figure}

\begin{figure}[hbt!]
    \centering
    \includegraphics[width=0.45\linewidth]{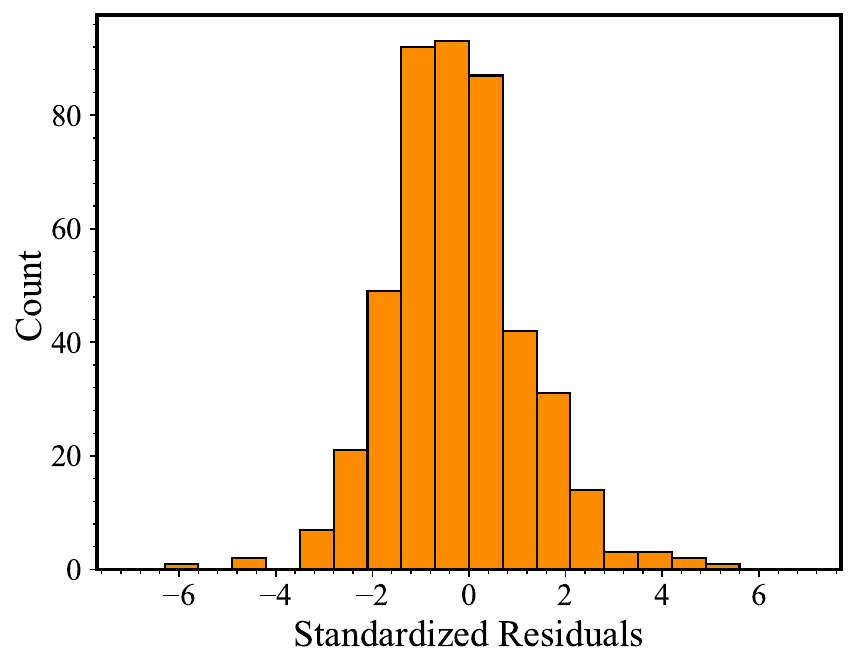}
    \includegraphics[width=0.45\linewidth]{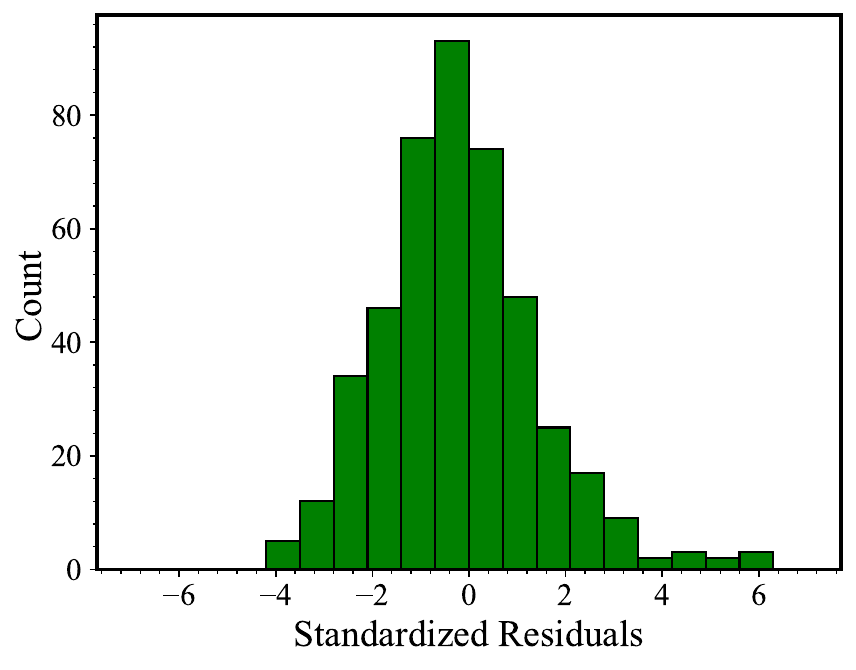}
    \caption{ The distribution of the standardized residuals for Models I (orange) and II (green). There are 20 bins in region [-7, 7].}  
    \label{SR20}
\end{figure}

\end{document}